\documentclass[sigconf]{acmart}

\usepackage{graphicx}
\usepackage[linesnumbered, ruled]{algorithm2e}
\usepackage{multirow}
\usepackage{makecell}
\usepackage{enumitem}
\usepackage{amsmath}
\usepackage{subfigure}
\usepackage{algorithmic}

\setcopyright{acmcopyright}
\copyrightyear{2023}
\acmYear{2023}
\setcopyright{acmlicensed}\acmConference[WWW '23]{Proceedings of the ACM Web Conference 2023}{May 1--5, 2023}{Austin, TX, USA}
\acmBooktitle{Proceedings of the ACM Web Conference 2023 (WWW '23), May 1--5, 2023, Austin, TX, USA}
\acmPrice{15.00}
\acmDOI{10.1145/3543507.3583374}
\acmISBN{978-1-4503-9416-1/23/04}

\begin{document}

\title{Robust Preference-Guided Denoising for\\ Graph based Social Recommendation}

\author{Yuhan Quan}
\affiliation{%
  \institution{Department of Electronic Engineering, Tsinghua University}
  \city{Beijing}
  \country{China}
  \postcode{100084}
}

\author{Jingtao Ding}
\authornote{Corresponding author (dingjt15@tsinghua.org.cn)}
\affiliation{%
  \institution{Department of Electronic Engineering, Tsinghua University}
  \city{Beijing}
  \country{China}
  \postcode{100084}
}

\author{Chen Gao}
\affiliation{%
  \institution{Department of Electronic Engineering, Tsinghua University}
  \city{Beijing}
  \country{China}
  \postcode{100084}
}

\author{Lingling Yi}
\affiliation{%
  \institution{Wechat, Tencent}
  \city{Shenzhen}
  \country{China}
  \postcode{518057}
}

\author{Depeng Jin}
\affiliation{%
  \institution{Department of Electronic Engineering, Tsinghua University}
  \city{Beijing}
  \country{China}
  \postcode{100084}
}

\author{Yong Li}
\affiliation{%
  \institution{Department of Electronic Engineering, Tsinghua University}
  \city{Beijing}
  \country{China}
  \postcode{100084}
}

\begin{abstract}
Graph Neural Network~(GNN) based social recommendation models improve the prediction accuracy of user preference by leveraging GNN in exploiting preference similarity contained in social relations. 
However, in terms of both effectiveness and efficiency of recommendation, a large portion of social relations can be redundant or even noisy, \textit{e.g.}, it is quite normal that friends share no preference in a certain domain.
Existing models do not fully solve this problem of relation redundancy and noise, as they directly characterize social influence over the full social network.
In this paper, we instead propose to improve graph based social recommendation by only retaining the informative social relations to ensure an efficient and effective influence diffusion, \textit{i.e.}, graph denoising.
Our designed denoising method is preference-guided to model social relation confidence and benefits user preference learning in return by providing a denoised but more informative social graph for recommendation models. 
Moreover, to avoid interference of noisy social relations, it designs a self-correcting curriculum learning module and an adaptive denoising strategy, both favoring highly-confident samples.
Experimental results on three public datasets demonstrate its consistent capability of improving two state-of-the-art social recommendation models by robustly removing 10-40\% of original relations. We release the source code at https://github.com/tsinghua-fib-lab/Graph-Denoising-SocialRec.
\end{abstract}


\maketitle

\section{Introduction}

Nowadays, social recommendation has become an important scenario of personalized recommender systems, where both user-item interactions and user-user social connections are available in the platforms~\cite{tang2013social}. 
Previous studies on human behaviors have established two basic theories describing the impact of social relations between users: 1) socially-connected users tend to possess similar preferences, which is referred as \textit{social homophily}~\cite{mcpherson2001birds}, and 2) behavior of a user can be influenced by her friends, like co-purchasing the same item, which is referred as \textit{social influence}~\cite{marsden1993network}.
Above theories regarding the impacts of social relations on recommendation have been validated by recent applications in areas like advertising~\cite{SocialAd}, e-commerce~\cite{curty2011social,xu2019think} and etc.

To characterize the above social relation effects on user preference, researchers have developed many social recommendation methods, like achieving preference similarity with social regularization~\cite{ma2011recommender}, enhancing influence from trust relations~\cite{guo2015trustsvd}, and exploiting different relations with strong or weak ties~\cite{wang2016social}.
Recent progress of graph neural network~(GNN) in graph machine learning~(ML) has further boosted the development of graph based social recommendation~\cite{gao2022survey}, as messaging passing mechanism widely adopted in GNN is well suited for characterizing influence diffusion in social context~\cite{fan2019graph,wu2019neural,tao2022revisiting}.
Moreover, aided by self-supervised learning techniques, GNN based social recommendation~(short for GSocRec) models can extract more informative signals from massive observed social relations~\cite{yu2021self,yu2021socially}.

\begin{figure*}[t]
    \centering
    \subfigure[Empirical data statistics]{\includegraphics[width=.23\textwidth]{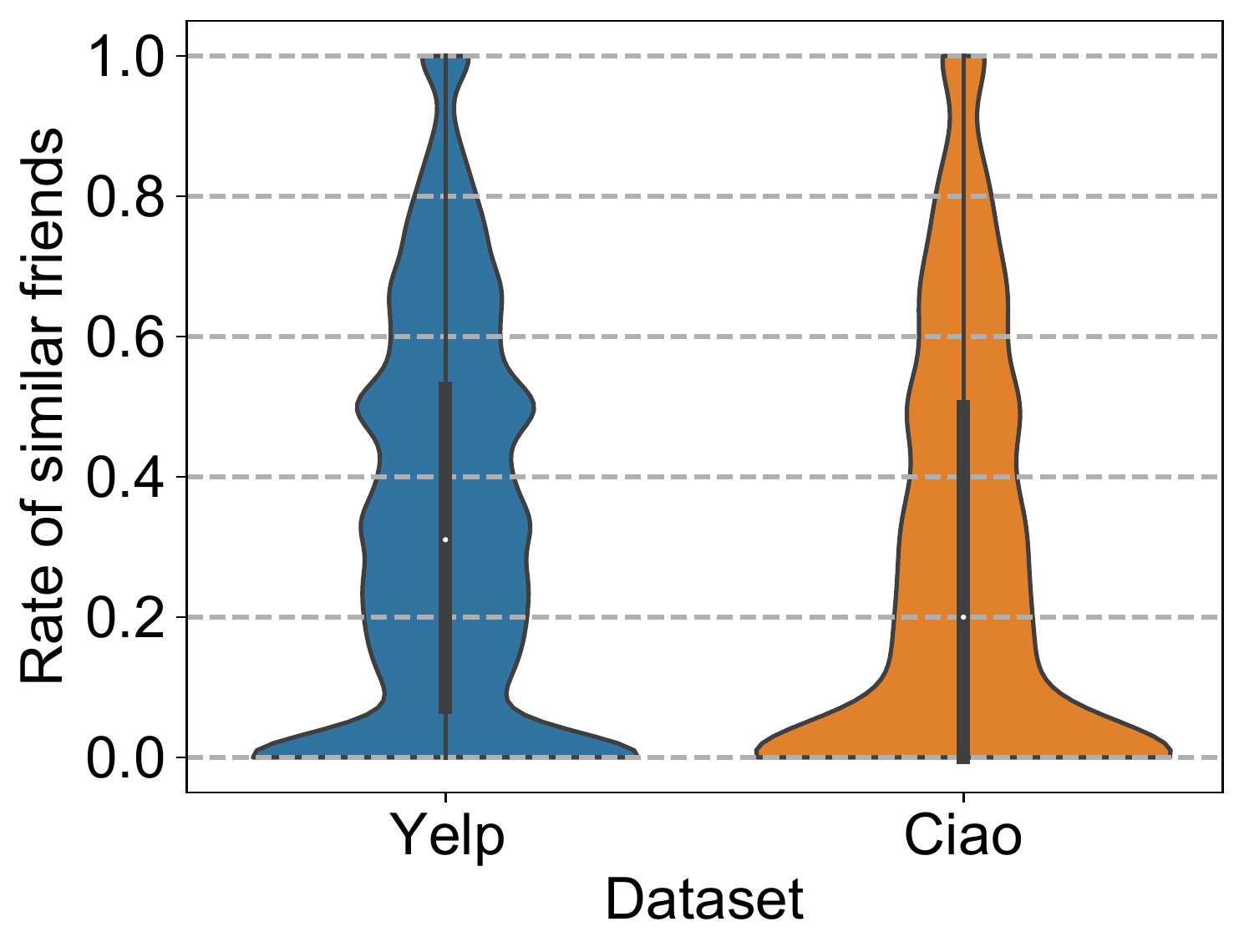}}
    \hspace{.1in}
    \subfigure[Proposed framework]{\includegraphics[width=.65\textwidth]{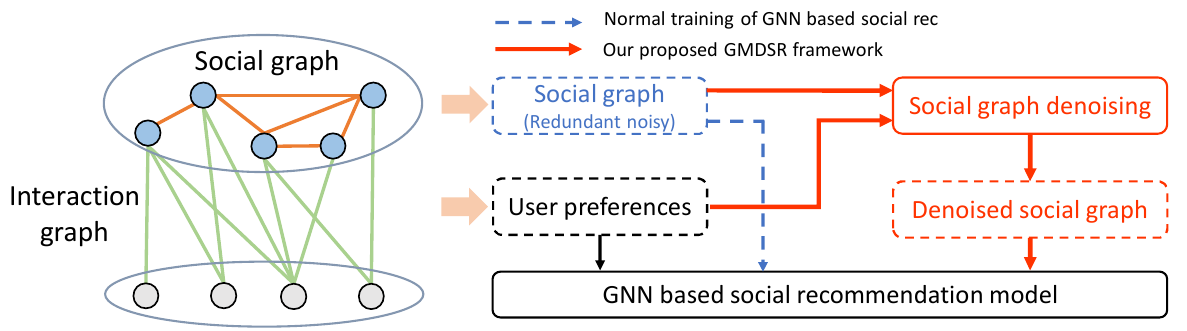}}
    \vspace{-4mm}
    \caption{(a) Distribution plot \textit{w.r.t.} ratio of friends having co-interactions. (b) Our proposed denoising enhanced social recommendation framework.}
    \label{fig:similar_friends_and_framework}
\end{figure*}

However, contrastive to the prevalence of GSocRec models, one important problem that has been barely explored is the existence of redundant or even noisy social relations in the context of graph based social recommendation. 
As we have shown in Figure~\ref{fig:similar_friends_and_framework}(a), most users only share similar preferences with a small portion of friends. Specifically, the median ratio value of friends having co-interactions is about 30\% and 20\% in two empirical datasets, respectively. Therefore, the rest social relations may be redundant or even harmful for social recommendation, as it is quite possible that two friends have distinct preferences in some areas.
Consequently, previous GSocRec models that directly characterize social influence over the full social network have two shortcomings. 
One is the huge burden on both computation and storage brought by learning GNN on the full user-user graph, as the size of social relations can easily approach to hundreds of billion in current commercial social recommendation platforms like Facebook and Wechat~\cite{lin2022platogl, sima2022ekko}~(both having over billions of users).  
This efficiency issue further hinders the graph structure learning based approaches~\cite{yu2020enhance,wei2022gsl4rec} that aim to learn a more informative graph by operating on the cartesian product space of nodes.
Another is the possible negative impact on recommendation accuracy. Although some previous works propose to deal with this diversity of social influence via attention mechanism~\cite{wu2020diffnet++,yang2021consisrec} or expectation-maximization method~\cite{wang2019social}, they still lack groundtruth labels and thus cannot learn the degree of social influence effectively.

To fully resolve the above problem, we propose to improve graph based social recommendation by denoising social network, \textit{i.e.}, removing those redundant or noisy social relations from the original graph and only retaining the informative ones for much more efficient and effective learning of GSocRec models. However, graph denoising for GSocRec models is non-trivial due to the following two challenges:
\begin{itemize}[leftmargin=*]
    \item How to overcome the difficulty of identifying the informative social relations for recommendation purpose? Reasons behind users' social relation formation are complicated, which do not necessarily induce similar user preferences, increasing the difficulty of identifying useful relations.
    \item How to achieve robust graph denoising given the existence of noisy labels? In the context of social recommendation, a large portion of user relations are useless or even harmful for enhancing user preference learning. Thus they can be seen as noisy signals and pose robustness requirements on graph denoising.
\end{itemize}

In this paper, we propose a novel denoising-enhanced recommendation framework for GSocRec models as illustrated in Figure~\ref{fig:similar_friends_and_framework}(b). The core of this framework is a \textbf{G}raph \textbf{D}enoising \textbf{M}ethod for \textbf{S}ocial \textbf{R}ecommendation~(shorten as GDMSR), which leverages user preferences to identify informative social relations from massive candidates and retain them as the denoised social graph, empowering the downstream GSocRec model with an arbitrary structure.
To solve the first challenge of characterizing relation confidence, GDMSR is designed to better exploit preference guidance in a two-fold manner, \text{i.e.}, explicit preference-based relation modeling and implicit co-supervision by a recommendation loss. 
As for the second challenge of robust denoising, it is equipped with a self-correcting curriculum learning module and an adaptive denoising strategy that both favor highly-confident samples~(\textit{i.e.}, useful social relations for preference learning).
To summarize, our main contributions are as follows.
\begin{itemize}[itemsep=2pt,topsep=0pt,parsep=0pt,leftmargin=*]
    \item We rethink the reliability of social network in context of graph based social recommendation and provide a novel angle of improving by graph denoising.
    \item  We design a graph denoising framework that is preference-guided to model social relation confidence and benefits user preference learning in return, compatible with general GSocRec models.
    \item Experiment results on three public datasets demonstrate the superiority of our GDMSR over state-of-the-art baselines in terms of empowering a series of GSocRec models. Ablation studies on both empirical and synthetic datasets further support the rationality behind specific method design. Moreover, aided by a robust preference-guided graph denoising capability, GDMSR can achieve 10-40\% of reduction ratio on social graphs without hurting recommendation accuracy, as well as maintaining high practicality in terms of efficiency and scalability.
\end{itemize}

\section{Preliminaries}\label{sec:background} 
\subsection{GNN based Social Recommendation Model} 
Generally, a typical GNN based social recommendation~(short for GSocRec) model involves two key parts of social influence diffusion and user preference learning. Specifically, the former characterizes the impact of friends' preferences on a certain user, while the latter stands for training a prediction model based on user-item interactions, as a common step in recommender systems.
Mathematically, a social network of $N$ users is denoted as a user-user graph $\mathcal{G}_s(\mathcal{U}, \mathcal{E})$, where $\mathcal{U}=\{u\}$ and $\mathcal{R}=\{(u,v)|r_{uv}=1, \forall u,v \in \mathcal{U} \}$ denote the set of users and social relations, respectively. Besides, there is another bipartite graph $\mathcal{G}_r(\{\mathcal{U}\cup \mathcal{I}\}, \mathcal{P})$ that stands for the interactions $\mathcal{P}$ between $N$ users $\mathcal{U}$ and $M$ items $\mathcal{I}$.
To represent users and items, GSocRec models normally use $F$-dimensional latent embedding matrices, \textit{i.e.}, $\textbf{E}_1 \in \mathbb{R}^{N\times D}$ and $\textbf{E}_2 \in \mathbb{R}^{M\times D}$, that can be learned from one-hot encoding of ids~(users or items), continuous feature values, discrete feature ids, or their concatenations.

To disentangle social influence and user's own preferences, GSocRec models further use two different user preference representations, \textit{i.e.}, $\textbf{E}_{1,s}$ and $\textbf{E}_{1,r}$, to characterize above two types of effects, respectively.  
For social influence diffusion, GSocRec models leverage the message-passing mechanism of GNN to model the influence diffusion of a user $u$'s friends, denoted as $\mathcal{R}_u=\{v|r_{uv}=1\}$, on $u$'s preference. Specifically, the $u$'s preference representation after $K$-hop social influence propagation is 
\begin{equation}
\textbf{E}^{(K)}_{1,s}(u) = \text{GNN}\left(\textbf{E}^{(K-1)}_{1}(u), \left\{\textbf{E}^{(K-1)}_{1}(v)|\forall v \in \mathcal{R}_u\right\}\right),
\label{social_effect}
\end{equation}
where the $\mathrm{GNN}$ module can be an arbitrary GNN model such as GCN~\cite{kipf2016semi}, GAT~\cite{velivckovic2017graph}, GraphSAGE~\cite{hamilton2017inductive} and etc. 
On the other hand, given $u$'s interaction history $\mathcal{P}_u=\{i|p_{ui}=1\}$, \textit{i.e.}, the set of items that $u$ has interacted before, the preference representation of $u$ after $K$-hop preference propagation is
\begin{equation}
\textbf{E}^{(K)}_{1,r}(u) = \mathrm{GNN}\left(\textbf{E}^{(K-1)}_{1}(u), \left\{\textbf{E}^{(K-1)}_{2}(i)|\forall i \in \mathcal{P}_u\right\}\right),
\end{equation}
where another $\mathrm{GNN}$ module is used to update $\textbf{E}^{(K)}_{1,r}$.
The comprehensive representation of user preferences after considering both social influence diffusion and user preference learning is
\begin{equation}
\textbf{E}^{(K)}_{1}(u) = \mathrm{Combine}\left(\textbf{E}^{(K)}_{1,s}(u), \textbf{E}^{(K)}_{1,r}(u)\right),
\end{equation}
where ${E}^{(0)}_{1}(u)={E}_{1}(u)$ and the $\mathrm{Combine}$ module can be either a mean pooling operation~\cite{fan2019graph,tao2022revisiting} or an attentive aggregation operation~\cite{wu2020diffnet++}. 
As for item representations, it is also calculated in a similar message-passing manner as 
\begin{equation}
\textbf{E}^{(K)}_{2}(i) = \mathrm{GNN}\left(\textbf{E}^{(K-1)}_{2}(i), \left\{\textbf{E}^{(K-1)}_{1}(u)|\forall u \in \mathcal{P}_i\right\}\right),
\label{item_rep}
\end{equation}
which aggregates information propagated from item $i$'s neighbors $\mathcal{P}_i=\{u|p_{ui}=1\}$, \textit{i.e.}, the set of users that have interacted with $i$.

Then, based on above $2\times(K+1)$ representations of $u$ and $i$, the prediction score of $u$'s preference over $i$, denoted as $\hat{p}_{ui}$, is calculated as

\begin{equation}
\begin{aligned}
&\hat{p}_{ui} = \textbf{E}^{*}_{1}(u) \cdot \textbf{E}^{*}_{2}(i),\\
\text{where }\textbf{E}^{*}_{1}(u) =& \frac{\sum_{k=0}^K \textbf{E}^{(k)}_{1}(u)}{K+1},  \textbf{E}^{*}_{2}(i) = \frac{\sum_{k=0}^K \textbf{E}^{(k)}_{2}(i)}{K+1}.
\end{aligned}
\end{equation}

To train a recommendation model, a widely-adopted approach is to minimize following Bayesian Personalized Ranking~(BPR) loss function~\cite{BPR}:
\begin{equation}
\mathcal{L}^{BPR} = \sum\nolimits_{(u,i,j)\in\overline{\mathcal{P}}} -\ln{\sigma(\hat{p}_{ui}-\hat{p}_{uj})},
\label{L_bpr}
\end{equation}
where $\sigma$ is a sigmoid activation function and $\overline{\mathcal{P}}=\{(u,i,j)|(u,i)\in \mathcal{P}, (u,j)\notin \mathcal{P}\}$ contains training triples $(u,i,j)$ made up of an observed user-item interaction and another unobserved one.
Learning above BPR objective is equivalent to maximizing the likelihood of observing such pairwise ranking relations $\hat{p}_{ui} > \hat{p}_{uj}$.

\subsection{Discussion on Shortcomings} 
Previous works on GSocRec models generally follow the above paradigm, which focuses on enhancing user preference learning by leveraging the effect of social influence. Therefore, the vital part of GSocRec models is the modeling of social influence diffusion that aims to identify friends with similar interests as the target user and propagate their influence to this user effectively. 
To achieve this, GSocRec models characterize diversity of social influence by an attention mechanism in message passing, \textit{i.e.}, using GAT in Equation~\eqref{social_effect}~\cite{wu2019dual,wu2020diffnet++,yang2021consisrec}.
However, all these methods have to learn a recommendation model on a full $\mathcal{G}_s$, meaning that massive but useless message-passing operations are conducted between those weakly-connected users, which causes a severe efficiency issue.
Moreover, since social relations between users are formed because of various reasons, it is quite possible that $u$ and $v$ share no interest or have contrast interests in a certain domain. Thus retaining $(u,v)$ can even degrade recommendation accuracy, as it introduces noisy information. In this circumstance, besides the redundancy issue, the noise problem should also be tackled in GSocRec models.

\section{Proposed Method}

\begin{figure*}
    \centering
    \includegraphics[width=.96\textwidth]{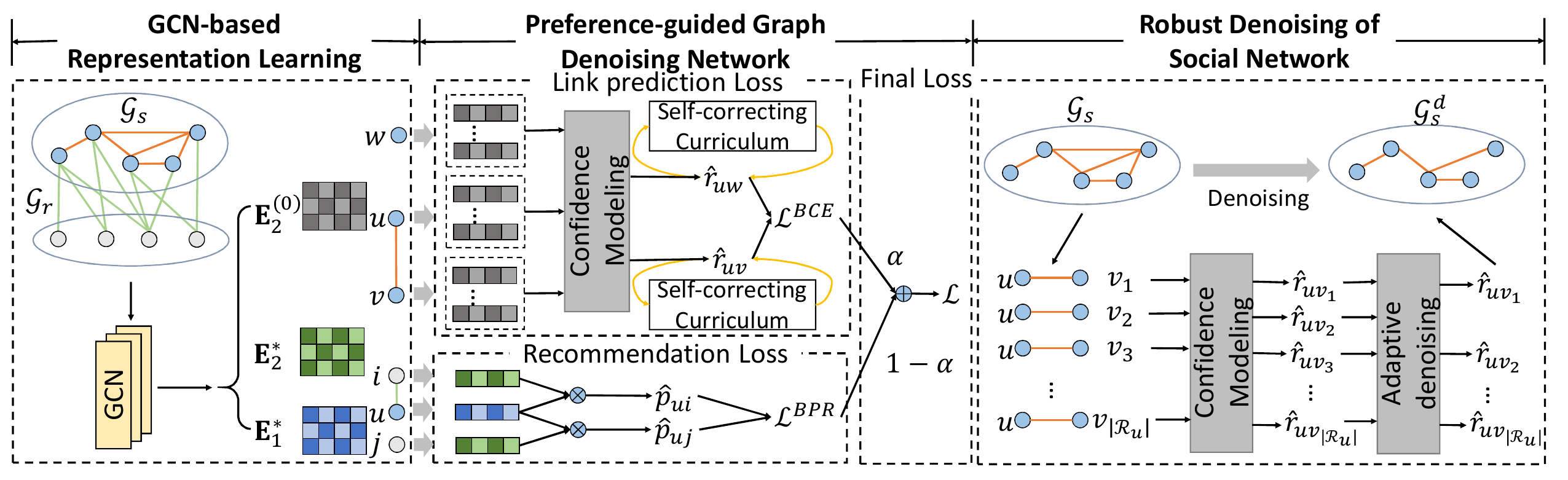}
    \vspace{-4mm}
    \caption{Details of graph denoising process in GDMSR.}
    \label{fig:model_detail}

\end{figure*}

To resolve the problems brought by redundant and noisy social relations that have hindered effective learning of social recommendation models, we propose the GDMSR method that can serve as a general framework for enhancing an arbitrary GNN based social recommendation model by denoising the original social graph. On the one hand, GDMSR leverages user preference signals to guide the graph denoising process, which helps identify informative social relations in the context of social recommendation. On the other hand, to achieve robust learning on the original social graph with noisy relations, GDMSR further incorporates a self-correcting curriculum learning mechanism that is less impacted by these noisy labeled data.
After the above effective and robust graph denoising training process, the obtained model is used to denoise the social network in an adaptive manner.

\subsection{Denoising-enhanced Social Recommendation Framework}
Instead of directly training a social recommendation model given the whole social network $\mathcal{G}_s$, our proposed GDMSR first learns to denoise $\mathcal{G}_s$ and obtain a much smaller but more informative graph $\mathcal{G}^d_s$. Then a GNN based social recommendation model utilizes $\mathcal{G}^d_s$ as its social graph to enhance user preference learning from the interaction graph $\mathcal{G}_r$.

\textbf{Denoising user relations for social recommendation.} Given an original user-connected graph that contains redundant or noisy user relations~(\textit{i.e.}, edges), the denoising task is defined as a link prediction alike task~\cite{liben2007link} that aims to generate a confidence score $\hat{r}_{uv}$ of link formation for each user pair $(u,v)$. Differently, only those observed user relations are checked in the denoising process by ranking according to $\{\hat{r}_{uv}\}$. Finally, the denoised graph $\mathcal{G}^d_s$ is obtained by removing those $(u,v)$s with a rather lower $\hat{r}_{uv}$.

\textbf{Training a social recommendation model with the denoised graph.} Given $\mathcal{G}^d_s$, as well as $\mathcal{G}_r$, any social recommendation model can be incorporated into GDMSR with no limitation on its structure. Moreover, since two training stages of denoising and recommendation are decoupled, one specific $\mathcal{G}^d_s$ can cooperate with multiple types of recommendation models in a unified way.

Since GDMSR is generally workable for all GNN based social recommendation models discussed in Sec.~\ref{sec:background}, we focus on introducing the design of graph denoising network in the rest of this section, which has also been illustrated in Figure~\ref{fig:model_detail}.

\subsection{Preference-guided Graph Denoising Network} %
A key challenge of denoising user-connected social network for recommendation is the lack of groundtruth labels on redundant social relations, and thus every existed link is equally treated when training the model, making it infeasible to identify informative links from massive data.
Current social recommendation models expect to automatically capture the diverse social influence by using a learnable message-passing mechanism~(like GAT), which can be ineffective because of indirect supervision from recommendation loss.
Since the purpose of denoising is to obtain a sparsified but more informative social graph for user preference learning, we propose to design a preference-guided graph denoising network for solving the social relation redundancy issue in the context of social recommendation.

\textbf{GCN-based representation learning.}
Similar to GNN based social recommendation models, our designed preference-guided graph denoising network in GDMSR also adopts a GNN based model structure, with the same input of both social graph $\mathcal{G}_s$ and interaction graph $\mathcal{G}_r$.
Specifically, the user and item representations after $K$-hop propagation, \text{i.e.}, $\{\textbf{E}^{(k)}_{1}\}^{K}_{k=0}$ and $\{\textbf{E}^{(k)}_{2}\}^{K}_{k=0}$, are calculated as in Equation~\eqref{social_effect} to \eqref{item_rep}. 
Since we aim to improve the previous learning approach via attention mechanism, the specific GNN module is GCN instead of GAT.

\textbf{Link prediction training.}
Given a prediction of the confidence score between two users $u$ and $v$, \textit{i.e.}, $\hat{r}_{uv}$, the link prediction problem aims to minimize the following binary cross-entropy~(BCE) alike loss function:
\begin{equation}
    \mathcal{L}^{BCE} \!=\! -\!\sum\nolimits_{(u,v)\in \mathcal{R}}\text{log}(\sigma(\hat{r}_{uv})) - \! \sum\nolimits_{(u,w)\notin \mathcal{R}}\text{log}(1-\sigma(\hat{r}_{uw})).
\end{equation}

In terms of leveraging user preference as the guiding signal for social graph denoising, GDMSR has the following two designs to better exploit the above useful knowledge.

\textbf{Preference-based relation confidence modeling.}
A common choice for calculating relation confidence $\hat{r}_{uv}$ is to design a scoring function $\phi(\cdot)$, \textit{i.e.}, 
\begin{equation}
\hat{r}_{uv}=\phi\left(\{\textbf{E}^{(k)}_{1}(u)\}^{K}_{k=0}, \{\textbf{E}^{(k)}_{1}(v)\}^{K}_{k=0}\right),
\end{equation}
where $\phi(\cdot)$ can either be a learnable NN~(like MLP) or an non-parametric operation~(like inner product). 
However, we argue that this way of relation confidence modeling is not precise, as the inputs $\{\textbf{E}^{(k)}_{1}\}^{K}_{k=0}$ may take in unrelated information. 
To find influential friends in the context of social recommendation, it is intuitive to directly compare their interaction history because these characterize the similarity of their preferences.
However, above GCN based user representations $\{\textbf{E}^{(k)}_{1}\}^{K}_{k=0}$ embeds a more comprehensive view of user traits, including user profile~(\textit{i.e.}, $\textbf{E}^{(0)}_{1}$), a fusion of user preference and social influence~(\textit{i.e.}, $\textbf{E}^{(1)}_{1}$) and higher order ones.
To ensure simple but effective guidance from user preferences, our proposed GDMSR only use the interaction history to characterize the relation confidence between two friends.
Mathematically, $\hat{r}_{uv}$ is calculated as follows,
\begin{equation}
\!\!\hat{r}_{uv}=\mathrm{Trf}\left(S_L\!\left(\left\{\textbf{E}^{(0)}_{2}(i)|\forall i\in \mathcal{P}_u\right\}\right) \oplus S_L\!\left(\left\{\textbf{E}^{(0)}_{2}(j)|\forall j\in \mathcal{P}_v\right\}\right)\right),
\label{relation_model}
\end{equation}
where a $\mathrm{Transformer}$ module~\cite{vaswani2017attention}, denoted as $\mathrm{Trf}$, is used given its power in  modeling similarity between two sequences of user interaction history.
Specifically, the input is the concatenation of two sequence embeddings~(denoted as $S_L(\cdot)$) representing the interaction history of $u$ and $v$, respectively, each with a fixed length of $L$ Since no positional information is required in modeling relation confidence, here we do not apply positional encoding in the Transformer.
As for the output, we add a ``CLS'' alike token at the end of the input sequence and use the corresponding transformer-encoded embedding vector to calculate $\hat{r}_{uv}$ with a $\mathrm{MLP}$.

\textbf{Co-optimization with recommendation loss.}
Besides designing the above preference-based relation confidence modeling structure, we further enhance preference guidance by co-optimizing the denoising model with a recommendation loss. 
As shown in Equation~\eqref{relation_model}, the quality of item embeddings is vital in characterizing relation confidence between users.
Motivated by the idea of improving representation quality with self-supervised learning, we propose to add a recommendation loss~($\mathcal{L}^{BPR}$ in Equation~\eqref{L_bpr}) and simultaneously train item embeddings to predict user preferences, which benefits their use in graph denoising in return.
Therefore, the final loss function is as follows,
\begin{equation}
\mathcal{L}=\alpha\mathcal{L}^{BCE}+(1-\alpha)\mathcal{L}^{BPR},
\end{equation}
where $\alpha\in[0,1]$ is a hyperparameter that controls the relative importance of link prediction loss and recommendation loss.

In a word, our proposed GDMSR can denoise social relations by exploiting user preference in a two-fold manner, \textit{i.e.}, explicit preference-based relation modeling and implicit co-supervision from preference learning, which both help solve the social relation redundancy issue.

\subsection{Robust Denoising of Social Network}

Another key challenge of denoising social network for recommendation is the existence of noisy social relations~(\text{i.e.}, friends with no shared preference.) that increases the learning difficulty of the denoising model. 
For example, suppose there are two connected users $\{u,v\}$ with distinct preferences, having two sets of interacted items $\{\mathcal{P}_u,\mathcal{P}_v\}$ that are less correlated with each other, pushing $\hat{r}_{uv}$ in Equation~\eqref{relation_model} to a large value is generally difficult and may induce a biased model that has to memorize this difficult sample~\cite{memorization}.
Therefore, the proposed GDMSR designs a self-correcting curriculum learning mechanism and an adaptive denoising strategy to alleviate this noisy effect and thus achieve robust graph denoising. 

\textbf{Self-correcting curriculum learning mechanism.} during the GDMSR learning process, relations with a high probability of being noisy, \textit{i.e.}, with a rather low confidence score predicted by the current denoising model, are removed, and the set of removed relations maintains a dynamic update with the model every several training epochs.
This equals following a difficulty-based curriculum that tends to favor the rather easier samples so as to achieve robust learning under noise and keep correcting this curriculum by replacing the more difficult samples with the easier ones according to the current model~\cite{zheng2020error}.
Specifically, for a user $u$, the last $\eta_u$ of her friends ordered by $\hat{r}_{uv}$ are considered as noisy relations to remove from training. This set is updated every $D$ epochs, where $D$ is a hyperparameter representing the length of each period in the above self-correcting curriculum.

\begin{table*}[t]
    \centering
    \caption{Basic information of datasets.}
    \vspace{-4mm}
    \begin{tabular}{ccccccc}
        \toprule
         Dataset & \#Users & \#Items & \#Interactions & \#Relations & Interaction Density & Relation Density \\
         \midrule
         Ciao & 7,355 & 17,867 & 140,628 & 111,679 & 0.11\% & 0.21\% \\ 
         Yelp & 32,827 & 59,972 & 598,121 & 964,510 & 0.03\% & 0.09\% \\
         Douban & 2,669 & 15,940 & 535,210 & 32,705 & 1.26\% & 0.46\% \\
         \bottomrule
    \end{tabular}
    \label{Dataset}
\end{table*}

\textbf{Adaptive denoising strategy.} As illustrated in Figure~\ref{fig:model_detail}, the trained denoising model is used in the final stage for social network denoising, \textit{i.e.}, predicting the confidence score of each existed social relation and removing those with low confidence.
On the one hand, since the training process with noisy labels may be unstable, the robustness of GDMSR is further enhanced by smoothing its prediction results along the training process~\cite{huang2020self,chen2021beyond}.
Specifically, $\hat{r}_{uv}$ at the end of $k$th curriculum period, \textit{i.e.}, $kD$th epoch, is smoothed by that of the last period.
Mathematically, $\hat{r}_{uv}$ is updated as follows,
\begin{equation}
\hat{r}_{uv}(t=kD) = \beta \cdot \hat{r}_{uv}(t=(k-1)D) + (1-\beta) \cdot \hat{r}_{uv}(t=kD),
\end{equation}
where a hyperparameter $\beta$ controls smoothness.
On the other hand, Motivated by \textit{Dunbar's number theory} suggesting an upper limit of a user's close friends~\cite{dunbar2010many}, we propose to adaptively denoise the user $u$'s social graph based on her friend number, \textit{i.e.}, $|\mathcal{R}_u|$.
Specifically, for each user $u$, $\eta_u$ of $u$'s friends are removed from the original graph $\mathcal{G}_s$, where we use the same hyperparameter $\eta_u$ as in the above curriculum design.
Mathematically, the denoising ratio $\eta_u$ is calculated as follows,
\begin{equation}
    \eta_u=
    \left \{
    \begin{array}{ll}
    0, & \text{if } |\mathcal{R}_u|<\epsilon, \\
    \left[\lfloor\mathrm{log}_{10}(|\mathcal{R}_u|)\rfloor\right]^\gamma \times R, & \text{else},
    \end{array}
     \right. 
\end{equation}
where $\epsilon$, $\gamma$ and $R$ are three hyperparameters.
The core idea behind this formula is that sparsely-connected users~($|\mathcal{R}_u|<\epsilon$) can retain all their relations while densely-connected users get to cut more relations in the denoised graph $\mathcal{G}^d_s$, which is more robust comparing uniform dropping regardless of friend number.

In a word, by following the above self-correcting curriculum during the training process and adopting an adaptive manner of denoising, our proposed GDMSR is able to robustly denoise social relations based on user preferences. So far, we have completed the entire denoising process, and the whole algorithm is shown in the Appendix~\ref{sec:appendix_model}.

\section{Experiments}
\subsection{Experiment Settings}
\textbf{Dataset.}
We conduct experiments on three open social recommendation datasets, including Ciao, Yelp and Douban. These datasets have been widely used in social recommendation related works~\cite{wu2020diffnet++, yu2021self}. More dataset preprocessing details are in the Appendix~\ref{sec:appendix_experiment}. The statistical information of each dataset is shown in Table~\ref{Dataset}.

\textbf{Baseline.}
Since our GDMSR is able to generate denoised social networks and adapt to arbitrary social recommendation models. For social recommendation model, we choose Diffnet++~\cite{wu2020diffnet++} and MHCN~\cite{yu2021self} as baselines.
For the denoising method, we compare GDMSR with Rule based approach, NeuralSparse~\cite{zheng2020robust} and ESRF~\cite{yu2020enhance}. 
Finally, we also add LightGCN~\cite{he2020lightgcn} as a baseline, which only uses interaction data without social relation data.
More about the baseline method is in the Appendix~\ref{sec:appendix_experiment}.

\textbf{Evaluation.}
To evaluate the performance of all methods, we use $Recall@K$ and $NDCG@K$ as metrics, where K=\{1,3\}. 
According to the suggestions of \cite{zhao2020revisiting}, we use \textit{real-plus-N}~\cite{bellogin2011precision,said2014comparative} to calculate the measures. For each user in the test set, we randomly sample 100 items that the user has not interacted with and rank them with the positive samples in the test set.

More implementation details are in the Appendix~\ref{sec:appendix_experiment}.

\subsection{Overall Performance}

\begin{table*}[t]
 \setlength\tabcolsep{5pt}
    \centering
    \caption{Overall performance of our proposed method on different recommendation methods.}
    \vspace{-4mm}
    \begin{tabular}{ccccccccccc}
        \toprule
        \multicolumn{2}{c}{Dataset} & \multicolumn{3}{c}{Ciao} & \multicolumn{3}{c}{Yelp} & \multicolumn{3}{c}{Douban} \\
        Basemodel & Method & R@1 & R@3 & N@3 & R@1 & R@3 & N@3 & R@1 & R@3 & N@3 \\ 
        \midrule
        LightGCN & - & 0.2298 & 0.0785 & 0.2071 & 0.5861 & 0.2774 & 0.5804 & 0.4321 & 0.1696 & 0.4156   \\  
        \midrule
        \multirow{5}*{Diffnet++} & w/o denoising & 0.2742 & 0.1109 & 0.2639 & 0.6031 & 0.3072 & 0.5897 & 0.5165 & 0.2156 & 0.4988 \\  
        & Rule based & 0.2860 & 0.1123 & 0.2677 & 0.6230 & 0.3228 & 0.5996 & 0.5358 & \underline{0.2489} & 0.5172 \\   
        & NeuralSparse & \underline{0.2869} & 0.1153 & 0.2734 & \underline{0.6383} & \underline{0.3289} & \underline{0.6054} & \underline{0.5470} & 0.2226 & 0.5102 \\  
        & ESRF & 0.2864 & \underline{0.1197} &\underline{0.2736} & 0.6184 & 0.3124 & 0.5958 & 0.5374 & 0.2393 & \underline{0.5194} \\  
        & GDMSR & \textbf{0.3020} & \textbf{0.1244} & \textbf{0.2821} & \textbf{0.6449} & \textbf{0.3291} & \textbf{0.6102} & \textbf{0.5614} & \textbf{0.2540} & \textbf{0.5297} \\  
        & $\Delta$ & 5.26\% & 3.93\% & 3.11\% & 1.03\%& 0.06\% & 0.79\% & 2.63\% & 2.05\% & 1.98\%  \\
        \midrule
        \multirow{5}*{MHCN} & w/o denoising & 0.2330 & 0.0884 & 0.2297 & 0.6991 & 0.3252 & \underline{0.6364} & 0.6198 & 0.3167 & 0.5933 \\  
        & Rule based & 0.2301 & 0.0916 & 0.2311 & 0.6966 & 0.3234 & 0.6347 & 0.6082 & \underline{0.3372} & 0.5980 \\   
        & NeuralSparse & 0.2461 & \underline{0.1034} & 0.2540 & \underline{0.7012} & 0.3288 & 0.6352 & \underline{0.6206} & 0.3349 & \underline{0.6011} \\ 
        & ESRF & \underline{0.2495} & 0.1028 &\underline{0.2568} & 0.6927 & \underline{0.3298} & 0.6344 & 0.6194 & 0.3244 & 0.5995 \\ 
        & GDMSR & \textbf{0.2618} & \textbf{0.1138} & \textbf{0.2632} & \textbf{0.7036} & \textbf{0.3405} & \textbf{0.6434} &  %
        \textbf{0.6396} & \textbf{0.3496} & \textbf{0.6137} \\ 
        & $\Delta$ & 4.93\% & 10.06\% & 2.50\% & 0.34\% & 3.24\% & 1.10\% & 3.06\% & 3.68\% & 2.10\% \\

         \bottomrule
    \end{tabular}
    \label{tab:overall_performance}
\end{table*}

We compare the performance of GDMSR with original social recommendation models without social network denoising and two denoising methods. The results are shown in Table~\ref{tab:overall_performance}. The overall average denoising ratio for all denoising methods ranges from about 5\% to 10\% depending on the hyperparameters. All results are average value after five repeat experiments. We performed paired t-test between the results of GDMSR and best-performed baseline. All improvements are statistically signiﬁcant for p < 0.01. It can be observed that GDMSR achieves the best performance in all experiments and the max improvement is up to 10\% on one dataset, \textit{i.e.}, Ciao. 

Further, we have the following findings. \textbf{First}, compared to using the original social network directly, after denoising the social network through different methods, the performance is improved in all experiments. This shows that in social recommendation, unreliable social relations in social networks are ubiquitous, and not all social relations reflect the homogeneity or social influence of users. Even a simple ruled based denoising method can improve the performance. 
\textbf{Second}, compared with NeuralSparse which uses both user interaction information and social information for denoising, GDMSR has better performance. This shows that in complicated social networks, it is more effective to use only user preference than to use social relations and user preference for denoising. 
\textbf{Finally}, the proposed GDMSR can achieve performance improvements on the basis of different models. For example, on the Douban dataset, GDMSR achieved performance improvement of more than 2\% compared to the best baseline, using both Diffnet++ and MHCN, which means that GDMSR can identify useless relations without relying on specific models and can be more widely adapted to different social recommendation models.

\subsection{Ablation Study}

\begin{figure*}
    \centering
    \subfigure[co-optimization weight]{\includegraphics[width=.23\textwidth]{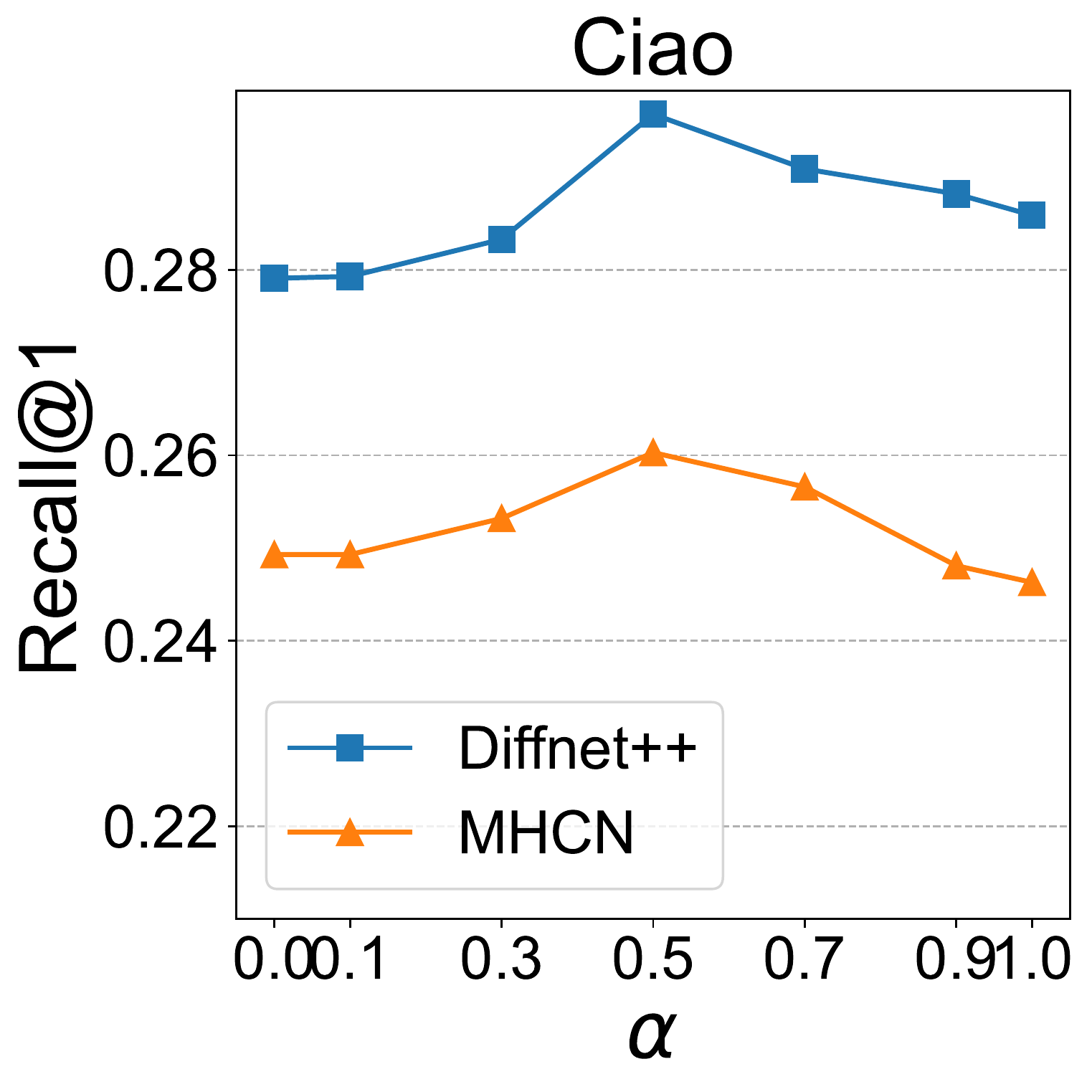}}
    \subfigure[co-optimization weight]{\includegraphics[width=.23\textwidth]{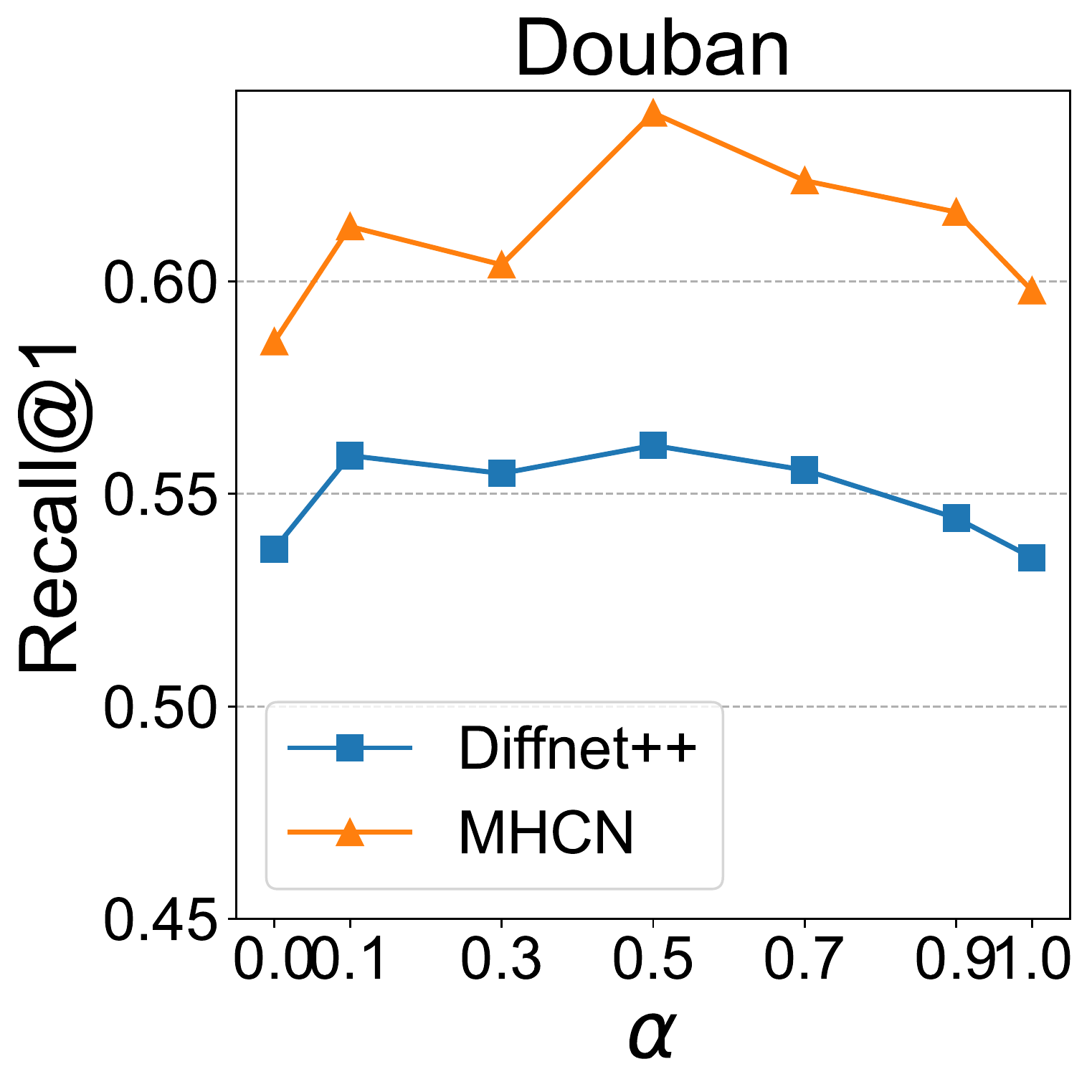}}
    \subfigure[relation confidence modeling]{\includegraphics[width=.235\textwidth]{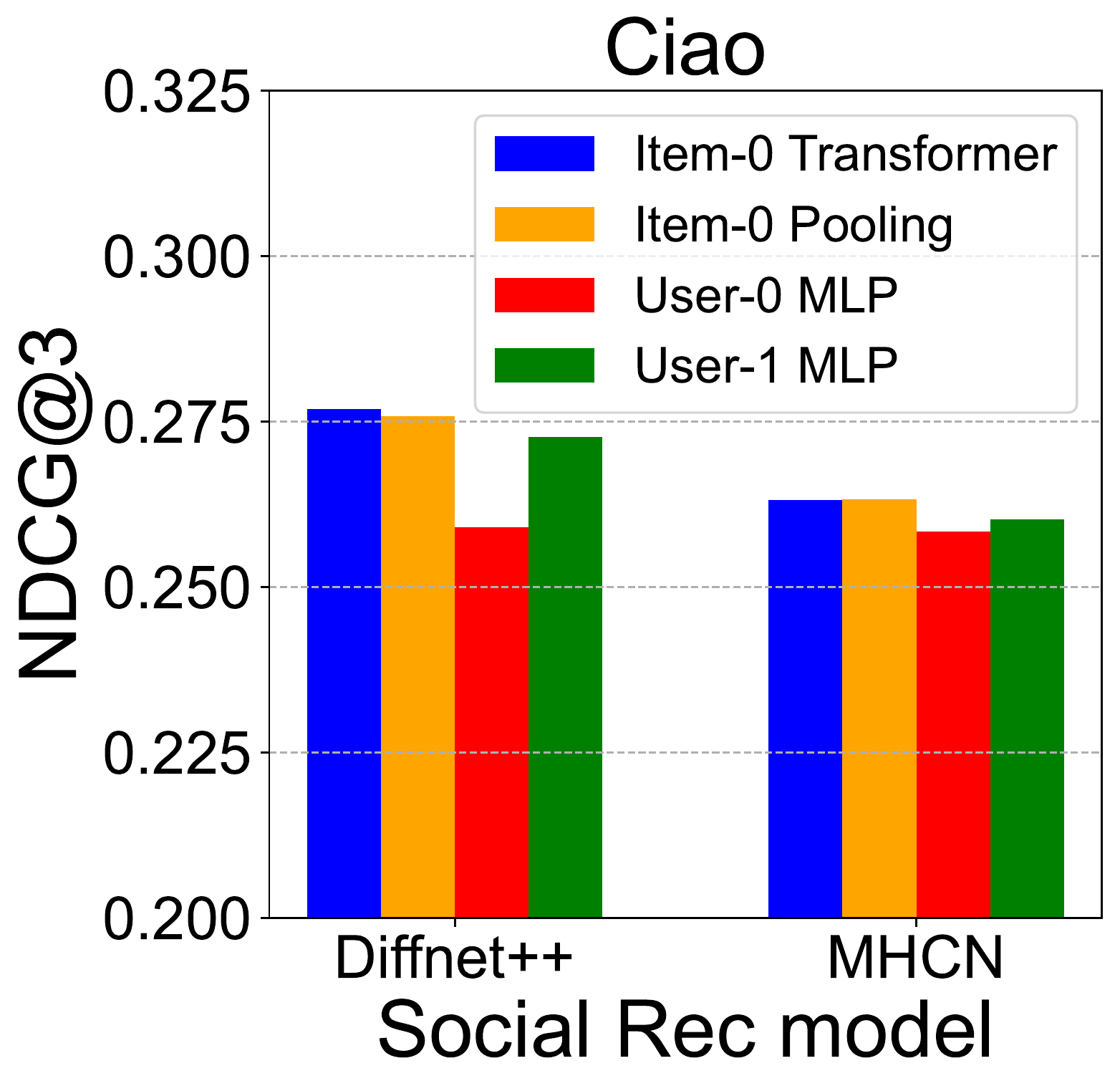}}
    \subfigure[relation confidence modeling]{\includegraphics[width=.23\textwidth]{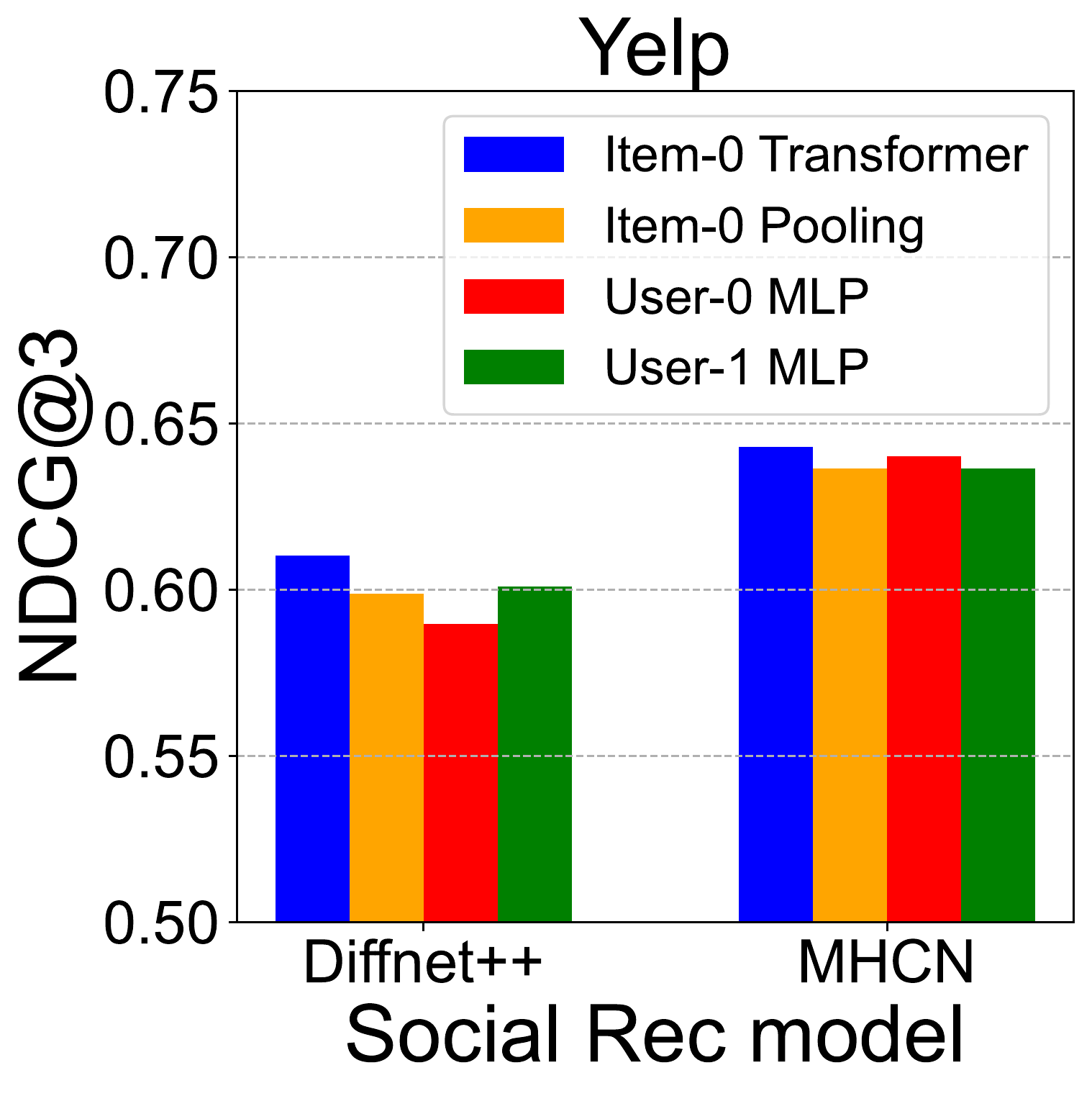}}
    
    \vspace{-4mm}
    \caption{(a)-(b) Performance on different $\alpha$ for co-optimization. (c)-(d) Performance comparison among different relation confidence modeling structures.}
    \label{fig:mlt_ablation_study}
\end{figure*}

\textbf{Preference-guidance from co-optimization.}
We propose to jointly optimize the recommendation loss and link prediction loss during denoising training, thereby improving the accuracy of relation confidence modeling. We tested the effect of co-optimization under different weights($\alpha$), and the results are shown in Figure~\ref{fig:mlt_ablation_study}(a)(b). It can be found that, on different social recommendation models, as $\alpha$ increases from 0 to 1, the performance first increases and then decreases. In detail, on the Ciao dataset, When $\alpha=0.5$, the three social recommendation models have achieved their respective best performance. On the Douban dataset, $\alpha=0.3$ or 0.5 works best. This proves that our co-optimization module can obtain better item representation through recommendation loss and further improve the accuracy of the link prediction task. At the same time, when $\alpha = 0$, the performance of the model has no obvious advantage compared with the method w/o denoising in Table~\ref{tab:overall_performance}, which shows that it is not enough to only model user preferences, and the denoising model needs the guidance of the link prediction task.

\begin{figure*}[t]
    \centering
    \subfigure[]{\includegraphics[width=.23\textwidth]{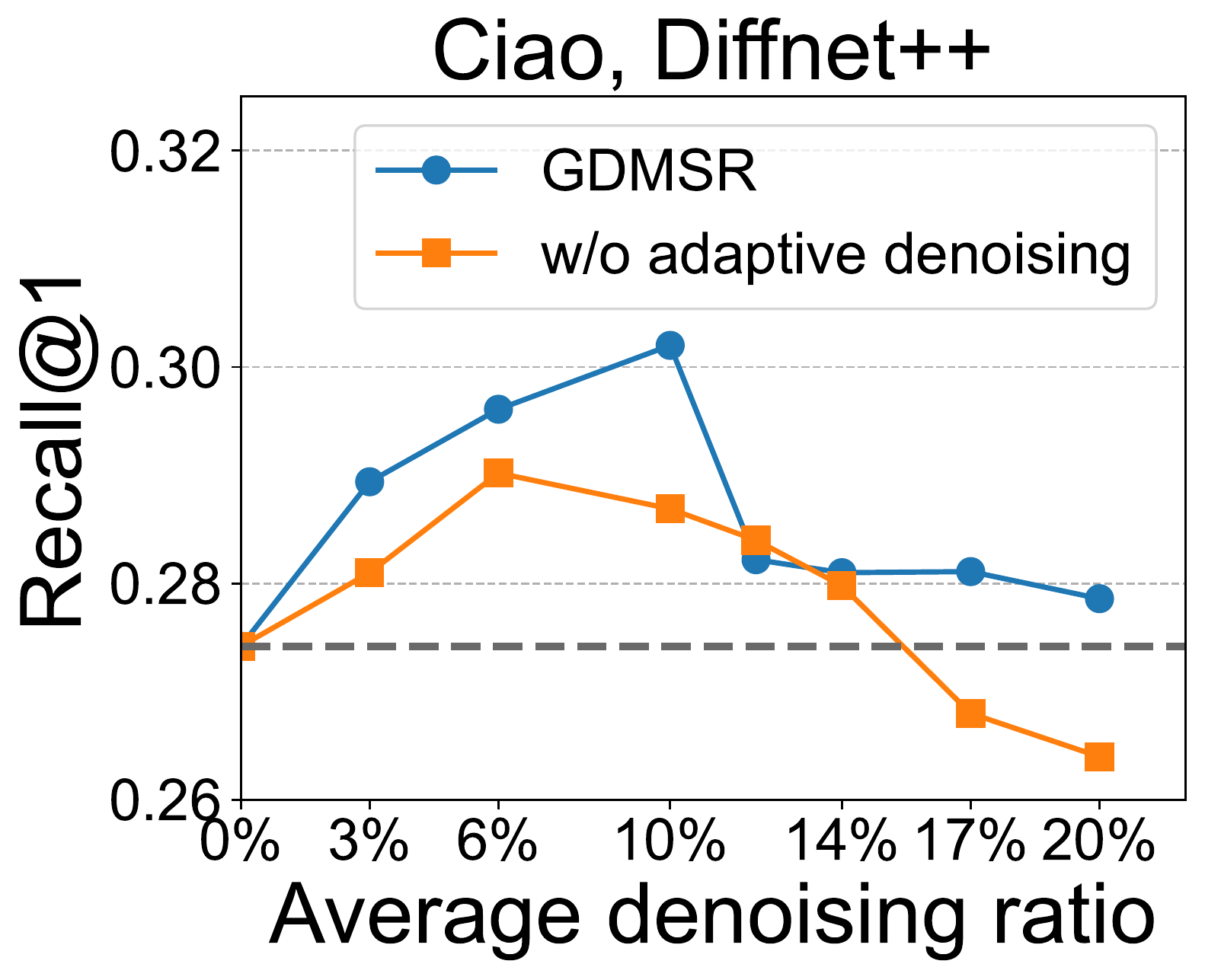}}
    \subfigure[]{\includegraphics[width=.23\textwidth]{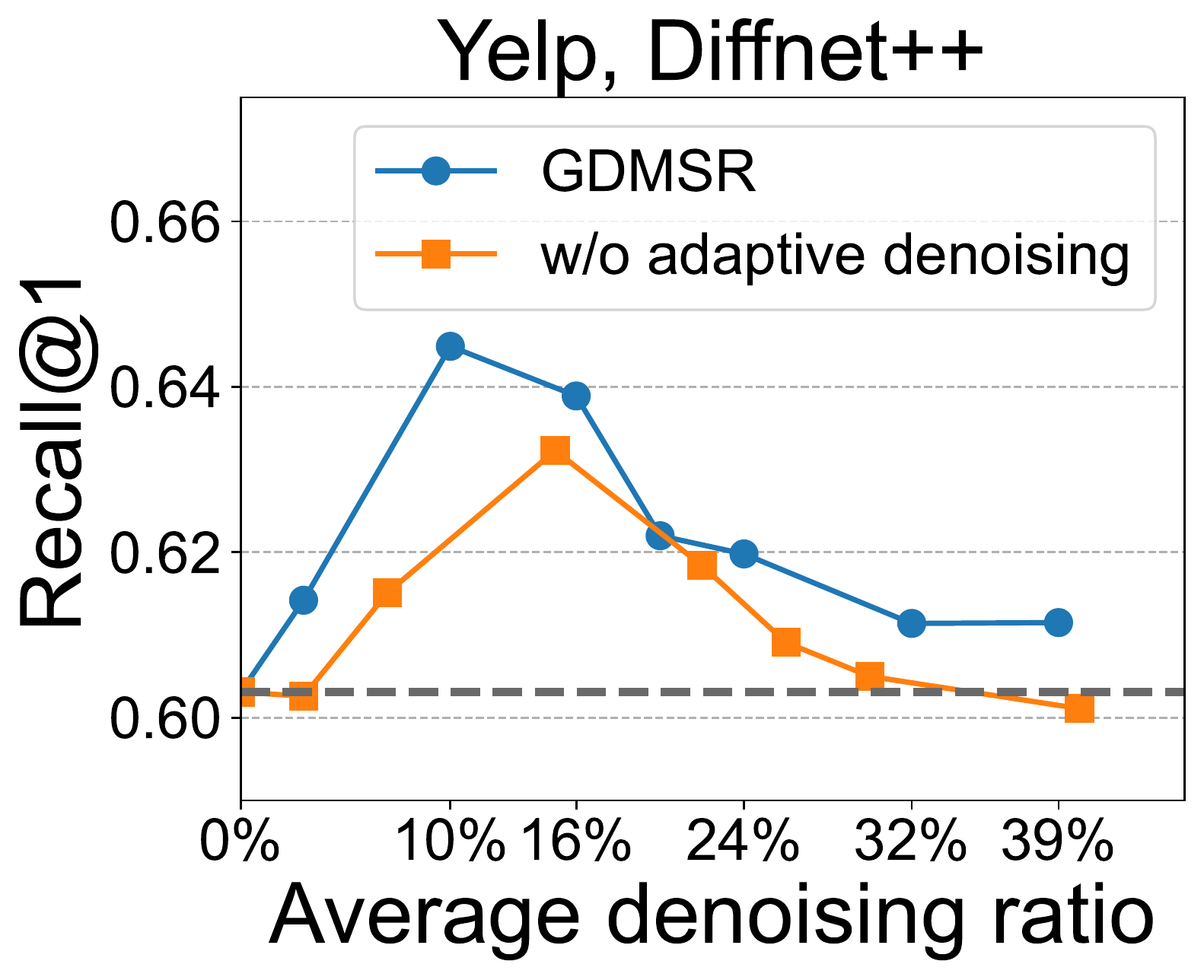}}
    \subfigure[]{\includegraphics[width=.23\textwidth]{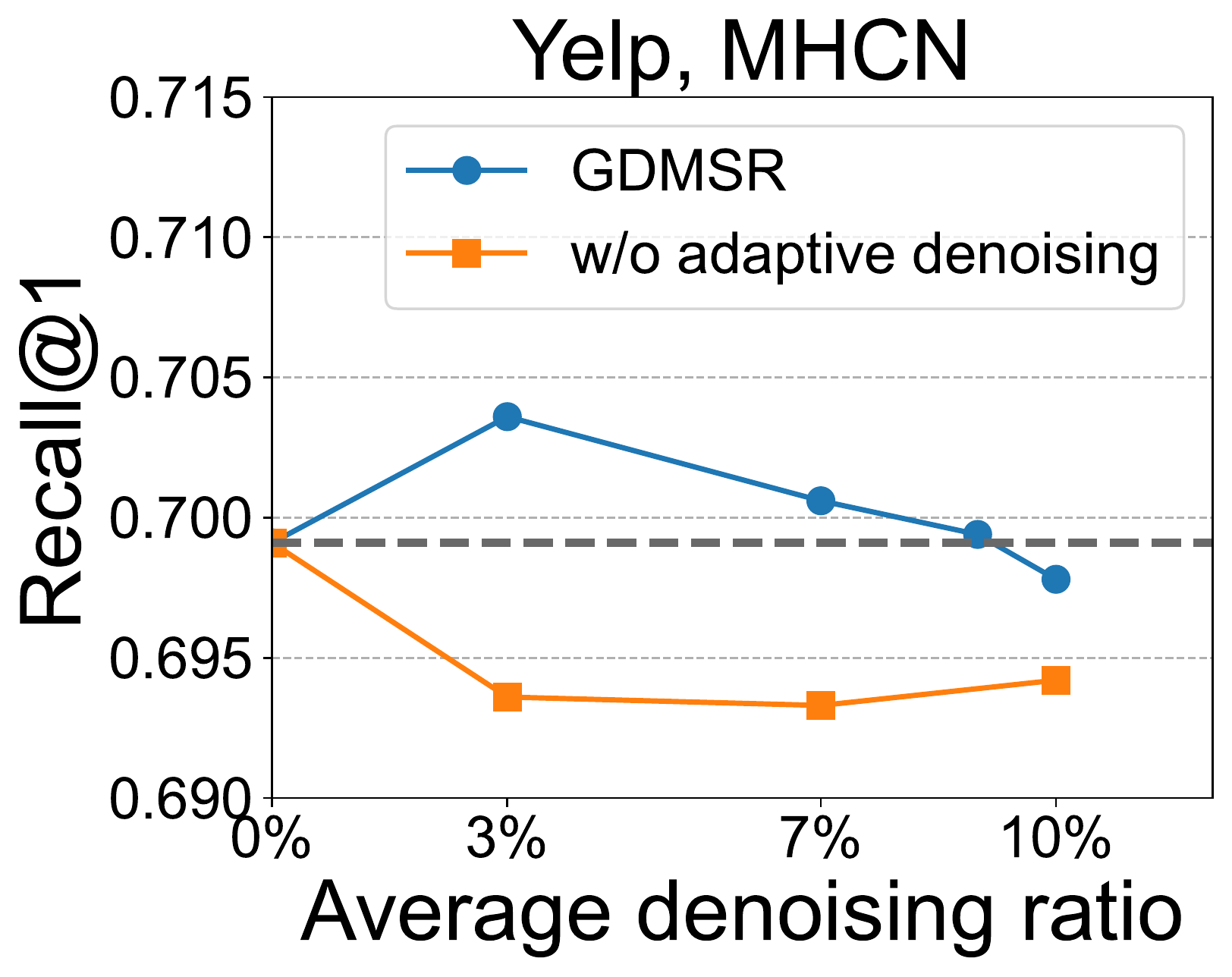}}
    \subfigure[]{\includegraphics[width=.23\textwidth]{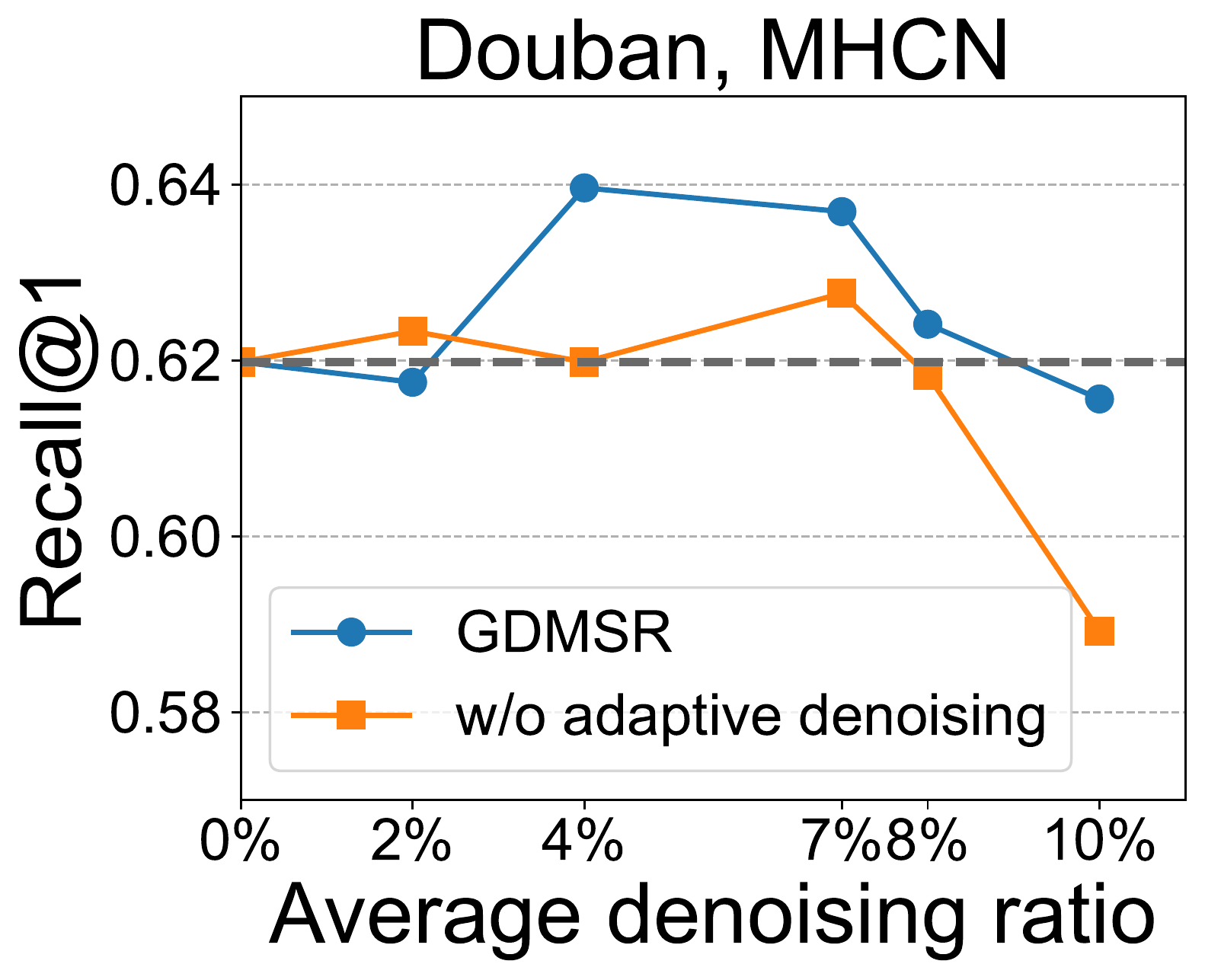}}
    
    \vspace{-4mm}
    \caption{Analysis of denoising robustness with respect to recommendation accuracy under different denoising ratio. }
    \label{fig:gamma_ablation_study}
\end{figure*}

\textbf{Modeling relation confidence with only preferences.}
In the relation modeling task, we propose to measure the confidence of the relation between users by only using the user's feedback, that is, the item representations that the user has interacted with. To verify the effect of this idea, we compare our proposed transformer-based relation confidence modeling module with other three options using different information, including using user representations or pooling the item representation directly. The results are shown in Figure~\ref{fig:mlt_ablation_study}(c)-(d). 
``User-0'' and ``User-1'' indicate using the representation of user nodes in GCN before convolution and from the first layer respectively.
We can observe that only using item representation works best, demonstrating the effectiveness of our proposed relation modeling structure, as it ensures simple but effective guidance from user preference~(interaction history). At the same time, compared to using user representation, only using item representation for pooling can achieve similar results. This again shows that the denoising model cannot effectively model the users' social relation confidence when integrating the user-side information including their profile~(\textit{i.e.}, $\textbf{E}^{(0)}_{1}$) and social influence~(\textit{i.e.}, $\textbf{E}^{(1)}_{1}$).

\textbf{Denoising robustness.}
In robust denoising module, we propose an adaptive denoising strategy, which makes the denoising ratio of users with more friends larger. To verify the effect of this strategy, we compare strategies that use the same denoising ratio~($\eta_u$) for all users. As shown in Figure~\ref{fig:gamma_ablation_study} and Appendix~\ref{sec:appendix_denoising}, the performance \textit{w.r.t.} $Recall@1$ in different cases consistently shows a trend of first increasing and then decreasing, indicating that proper denoising of social network can improve effectiveness of social influence, thereby improving the recommendation performance. Meanwhile, compared with non-adaptive denoising, the proposed GDMSR significantly outperforms in nearly all cases with different denoising ratio.
More importantly, our proposed GDMSR is able to achieve further sparsification without hurting original recommendation accuracy, which produce a memory-efficient social graph that is vital in practical applications.
Specifically, GDMSR combined with Diffnet++ is able to achieve 20-40\% of relation reduction while maintaining an accuracy improvement. Comparatively, this reduction value drops to about 10\% when combined with MHCN, which is still of large practical value in commercial recommender systems. Since MHCN relies on mining social motifs and leverages hypergraph convolution model to characterize their effects, graph denoising may bring larger impacts on MHCN because of structural changes.
In addition, we also compare the effect of choosing different value of hyperparameter $\gamma$. 
Due to space limit, please refer to Appendix~\ref{sec:appendix_denoising} for details.

\begin{figure}[t]
    \centering
    \subfigure[denoising capability]{\includegraphics[width=.25\textwidth]{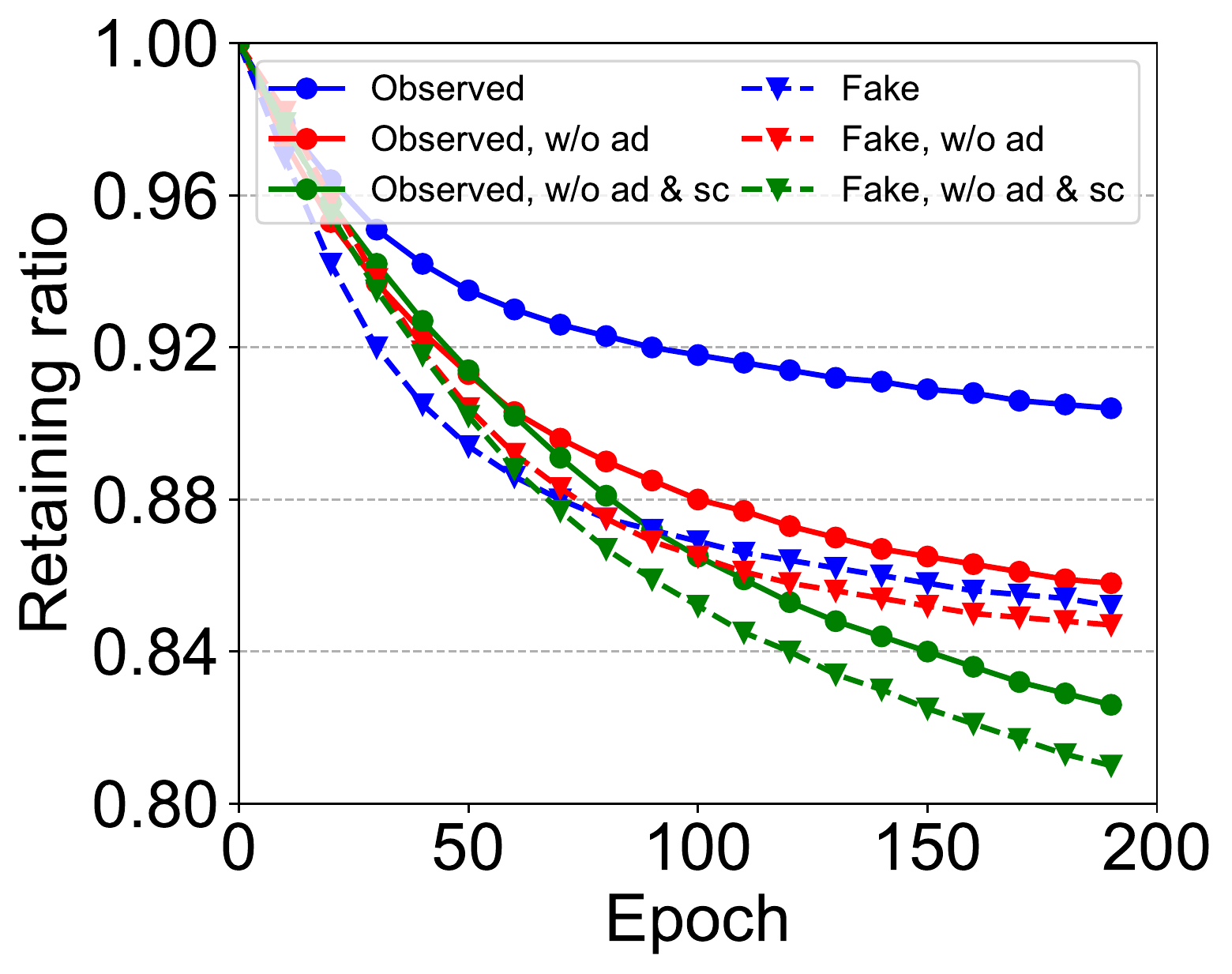}}
    \subfigure[inference efficiency]{\includegraphics[width=.22\textwidth]{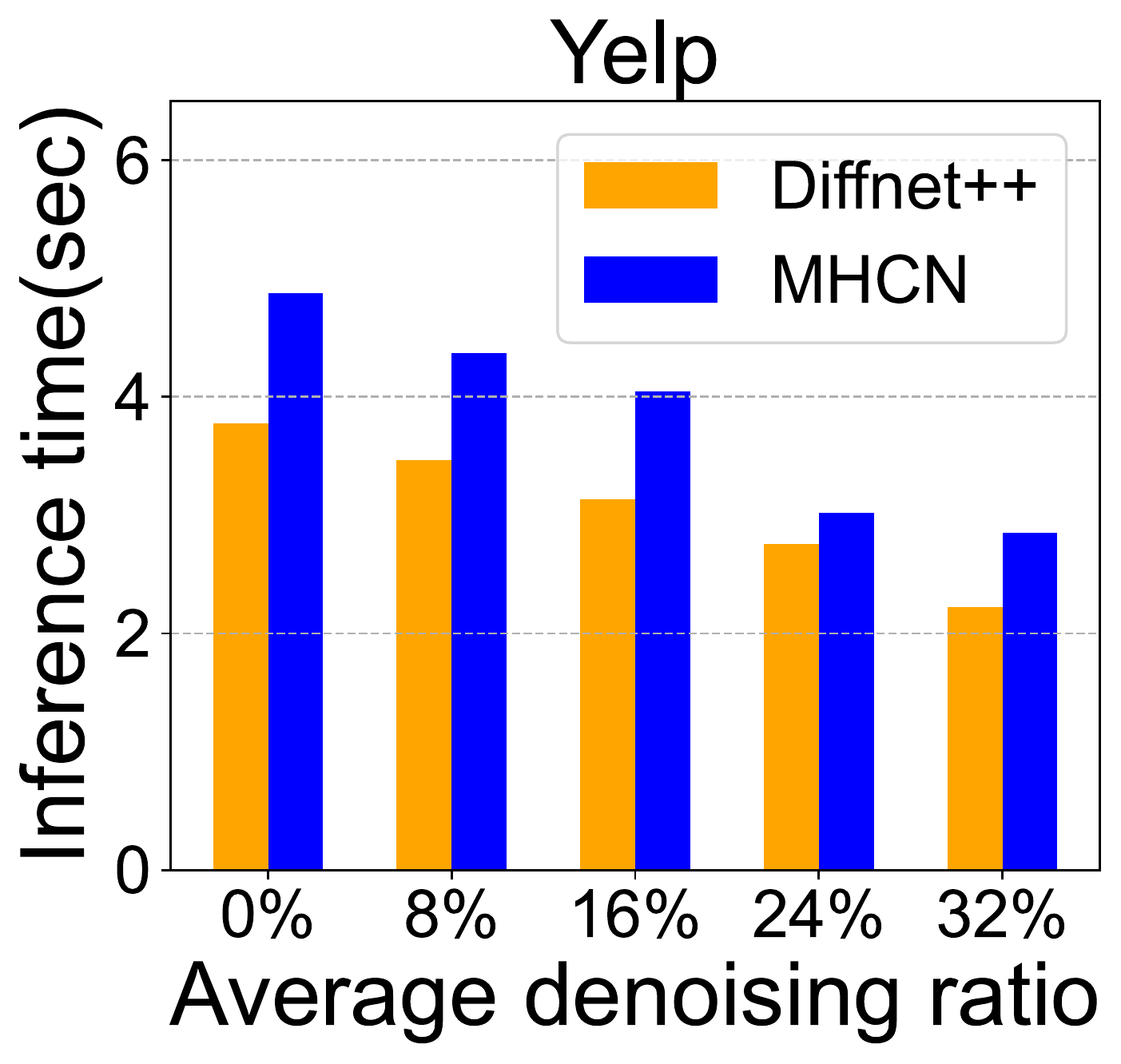}}
    \vspace{-5mm}
    \caption{(a) Retaining ratio of social relations adopting different denoising strategies~(synthetic). (b) Efficiency comparison
    \textit{w.r.t.} social graph related inference time~(Yelp).}
    \label{fig:syn_efficiency}
\end{figure}

\begin{table*}[t]
 \setlength\tabcolsep{5pt}
    \centering
    \caption{Overall performance of GDMSR based on 30\% of interaction data.}
    \vspace{-4mm}
    \begin{tabular}{ccccccccccc}
        \toprule
        \multicolumn{2}{c}{Dataset} & \multicolumn{3}{c}{Ciao} & \multicolumn{3}{c}{Yelp} &  \multicolumn{3}{c}{Douban} \\
        Basemodel & Method & R@1 & R@3 & N@3 & R@1 & R@3 & N@3 & R@1 & R@3 & N@3 \\ 
        \midrule
        \multirow{3}*{Diffnet++} & w/o denoising & 0.2742 & 0.1109 & 0.2639 & 0.6031 & 0.3072 & 0.5897  & 0.5165 & 0.2156 & 0.4988 \\  
        & GDMSR & \textbf{0.2899} & \textbf{0.1175} & \textbf{0.2755} & \textbf{0.6305} & \textbf{0.3189} & \textbf{0.6021}  & \textbf{0.5490} & \textbf{0.2497} & \textbf{0.5295}  \\  
        & $\Delta$ & 5.73\% & 5.95\% & 4.40\% & 4.54\% & 3.81\% & 2.10\% & 6.29\% & 15.82\% & 6.15\%  \\  
        \midrule
        \multirow{3}*{MHCN} & w/o denoising & 0.2330 & 0.0884 & 0.2297 & 0.6991 & 0.3252 & 0.6364  & \textbf{0.6198} & 0.3167 & 0.5933 \\  
        & GDMSR & \textbf{0.2616} & \textbf{0.1093} & \textbf{0.2638} & \textbf{0.7005} & \textbf{0.3438} & \textbf{0.6445} & 0.6148 & \textbf{0.3360} & \textbf{0.6012} \\  
        & $\Delta$ & 12.27\% & 23.64\% & 14.85\% & 0.20\% & 5.72\% & 1.27\% & -0.81\% & 6.09\% & 1.33\%  \\  
         \bottomrule
    \end{tabular}
    \label{tab:overall_performance_partial}
\end{table*}

\subsection{Denoising Capability on Synthetic Data}
To further evaluate denoising capability of our proposed GDMSR, we construct a synthetic dataset~(based on Ciao) with both observed social relations and fake social relations~(\textit{i.e.}, we randomly set some unconnected user pairs to be connected) and evaluate whether GDMSR is able to learn a powerful graph denoising model that can effectively discriminate a given relation label. 
Specifically, we inject the fake social relations with the same number of those observed ones, and further combine these corrupted relation data with original interaction data to train GDMSR. Then we record the retaining ratio of each part of relation data during the denoising process, and present the results in Figure~\ref{fig:syn_efficiency}(a). Here we also compare with two variants, \textit{i.e.}, without adative denoising~(shorten as ``w/o ad'') and without adaptive denoising and self-correcting curriculum learning~(shorten as ``w/o ad \& sc''). 
The results clearly indicate that GDMSR, even though under explicit noise of corrupted relation labels, can still maintain learning robustness by removing more fake relations~(16\%) than observed relations~(9\%). In other words, since we inject rather strong noise~(50\% of relation labels are corrupted), GDMSR manages to achieve an moderately good accuracy of 64\%~(16/(16+9)), while other two options cannot produce a workable denoising module under this level of label noise.

\subsection{Practicality Evaluation}
\textbf{Computation efficiency.} Our proposed GDMSR can produce a much sparser social graph~(\textit{e.g.}, 20-40\% smaller on Yelp), which can improve computation efficiency in inference period. In Figure~\ref{fig:syn_efficiency}(b) we present an inference time analysis of social graph related computations~(social-network-side GNN), which is conducted on Yelp dataset by inferring 50$w$ records. The results clearly demonstrate the increasing improvement of computation efficiency as denoising ratio goes larger. 

\textbf{Zero-shot denoising without frequent training.}
In industry practice, it is rather time consuming and nearly infeasible to frequently update graph denoising model given the massive social relation data~(up to hundreds of millions in Facebook and Wechat). Therefore, GDMSR is expected to support denoising training in a less frequent manner, \textit{i.e.}, produce high-quality distilled social graph without training on newly coming interaction data~\cite{lin2022platogl}.
To evaluate this practicality, we design another experiment by only using 30\% of interaction data to train denoising module of GDMSR and then use all interaction data as input to denoise social graph. This setting simulates the fact that GDMSR cannot be timely updated in practice. Instead, we may use the old GDMSR model but the new interaction data to directly denoise current social graph in a zero-shot-like manner.
Then, the denoised graph is used to train the social recommendation models and results are listed in Table~\ref{tab:overall_performance_partial}.
It can be observed that GDMSR is still able to effectively denoise the social network even when only 30\% of the training samples are used. In most cases, the recommendation accuracy can be improved compared to without denoising, with the highest improvement reaching 23\%, demonstrating that GDMSR has high scalability in terms of industrial practice.
In fact, this is mainly owing to our design of preference-guide denoising with only interactions, which is more robust and less influenced by other contextual changes.

\section{Related Work}

\subsection{Graph Denoising for Social Recommendation} 
Early works use simple statistics like number of co-interactions~\cite{ma2011recommender} to indicate the confidence degree of social relations, which is generally ineffective due to data sparsity~\cite{tang2013social}. 
In order to capture observed diversity of social influence, attention mechanism~\cite{wu2020diffnet++,yang2021consisrec}, fine-grained contextual information~\cite{fu2021dual}, expectation-maximization method~\cite{wang2019social} or reinforcement learning~\cite{du2022denoising} can be leveraged to learn adaptive weights among various friends. 
Although above techniques can be applied in prevalent GSocRec models, they generally fall short of learning the degree of social influence effectively due to a lack of groundtruth labels.
From a view of general graph machine learning, graph sparsification methods~\cite{zheng2020robust,dai2022towards} learn to drop edges from a general graph so as to achieve robust learning, which does not consider the specific social recommendation context. 
Moreover, graph structure learning methods~\cite{zhao2021heterogeneous,liu2022compact,liu2022towards} are optimized to not only drop but also add edges, which are too time-consuming for social recommendation as they generally operate on a huge space with cartesian product between users.
As a specific extension in social context, there are a series of works that propose to not only remove observed social relations with low quality but also add a few potential relations~\cite{yu2018adaptive,yu2019generating,yu2020enhance,krishnan2019modular}. This idea has also been adopted in dynamic recommendation scenarios like session recommendation by starting with observed social relations and updating relations during training process~\cite{song2020dream,wei2022gsl4rec}. In addition, the effectiveness of social recommendation can also be improved by using contextual features~\cite{krishnan2020transfer} or multi-view learning~\cite{krishnan2022multi}.
Different from above works, our proposed GDMSR method can produce a distilled social graph, not only sparser but also more informative, that is more suitable for real-world social recommender systems. In experiments we also compare with one competitive method above~\cite{yu2020enhance} to demonstrate effectiveness of GDMSR.

\subsection{Denoising for Recommendation} 
In general recommender systems, another trending topic is feedback denoising that enables more robust preference learning~\cite{he2021purify,tian2022learning}. Since there exists heterogeneity in users' manner of expressing preferences, it is non-optimal to set a unitary threshold for either explicit feedback or implicit feedback. For example, users differ in their criterion of rating items with low scores~\cite{wang2021denoising} or not clicking~\cite{ding2020simplify, ding2019reinforced, ding2019sampler}. There are also some works to solve this problem through causal inference~\cite{gao2022causal}.
To deal with this problem of learning from noisy labels~\cite{song2022learning}, existing works adopt techniques like adding graph-based priors~\cite{strahl2020scalable}, cross-model agreements~\cite{wang2022learning} and graph sparsification~\cite{gao2022self} to denoise observed user-item interaction data. 
Different from above works, we focus on denoising social network, \textit{i.e.}, user-user relation data, so as to leverage power of social homophily and social influence in a more effective and efficient way.

\section{Conclusion and Future Work}

In this paper, we approach the problem of graph based social recommendation by rethinking the reliability of the social network for learning GNN recommendation models.
Since there is no reliably labeled data, we choose to utilize user-item feedback that reflects user preference to filter noisy social relations, which, in turn, improves the preference learning.
Experiments on three public benchmark datasets and different social recommendation models not only show the effectiveness of the denoising method but also demonstrate our proposed solution can serve as a general framework. 
Precisely, the improvement is steady and significant, up to 10\%.

As for the future plan, one interesting direction is to deeply analyze the difference between the original social network and the denoised one, supported by the social network theory~\cite{centola2010spread}.

\clearpage

\begin{acks}
This work was supported in part by the National Key Research and Development Program of China under 2020AAA0106000, and the National Natural Science Foundation of China under 61971267, U1936217, 61972223, 62272262.

\end{acks}

\bibliographystyle{ACM-Reference-Format}
\bibliography{sample-base}

\clearpage
\appendix

\section{Training process of GDMSR}
\label{sec:appendix_model}

The overall process of GDMSR is shown in Algorithm~1.

\begin{algorithm}
\label{alg:GDMSR}
\caption{Training process of GDMSR}
    \begin{algorithmic}[1]
        \REQUIRE social relations set $\mathcal{R}$, interactions set $\mathcal{P}$
        \FOR{$epoch=1$ to MAX\_EPOCH}
            \STATE Generate training batch $\mathcal{B}$ from $\mathcal{P}$
            \FOR{$(u,i)$ in $\mathcal{B}$}
                \STATE sample $(u,j)$,$(u,v)$ and  $(u,w)$ with $j\notin \mathcal{P}_u$, $v\in \mathcal{R}_u$ and $w \notin \mathcal{R}_u$
                \STATE Calculate recommendation loss $\mathcal{L}^{BPR}$ by Eq.6
                \STATE Calculate link prediction loss $\mathcal{L}^{BCE}$ by Eq.7
                \STATE Calculate final loss $\mathcal{L}$ and update parameter by Eq.9
            \ENDFOR
            
            \IF{$epoch \% D == 0$}
                \FOR{$(u,v)$ in $\mathcal{R}$}
                    \STATE Calculate $\hat{r}_{uv}(t=kD)$ by Eq.8 and Eq.10
                \ENDFOR
                \FOR{$u$ in $\mathcal{U}$}
                    \STATE Calculate $\eta_u$ by Eq.11
                    \STATE Remove last $\eta_u$ of $u$'s friends ordered by $\hat{r}_{uv}(t=kD)$
                \ENDFOR
                    
            \ENDIF
        
        \ENDFOR
    
    \end{algorithmic}
\end{algorithm}

\section{Experiment Details}
\label{sec:appendix_experiment}

\subsection{Dataset preprocessing}
For all datasets, we only retain the samples with a score of 4 or 5 as positive samples. Besides, we filter users and items according to the number of interactions and the number of users' friends and only retain samples related to users and items with high activity. For the Yelp dataset, we retain the user attributes, and since the original data does not contain user attributes for the other two datasets, we only use the user id as a feature.

\subsection{Baselines}
Here is an introduction to the baseline methods we used

\textbf{Diffnet++}. This model uses graph attention networks to model diverse social influences among various friends.

\textbf{MHCN}. This model uses hypergraph convolution methods to capture different impacts of social motifs on preference learning among friends.

\textbf{Rule based approach}. Based on the social homogeneity assumption, the similarity between users can be measured according to the items that the users interact with. Therefore, we remove those relations with a few items that users have both interacted with. 

\textbf{NeuralSparse}. This is an end-to-end adaptive graph sparsification method that learns a differentiable sampler from empirical data by using Gumbel softmax. 

\textbf{ESRF}. This end-to-end social recommendation framework leverages Generative Adversarial Nets~(GAN) so as to generate more densely-connected social graphs for social recommendation.

\subsection{Running Environment}
The experiments are conducted on a single Linux server with AMD Ryzen Threadripper 2990WX@3.0GHz, 128G RAM and 3 NVIDIA GeForce RTX 2080TI-11GB. Our proposed GDMSR is implemented in PyTorch 1.10 and Python 3.7. Social recommendation models and baselines including SocialGCN, Diffnet++, MHCN and NeuralSparse are implemented in Tensorflow 1.14 and Python 3.7.

\subsection{Assets}
The code of our proposed GDMSR is included in the supplementary materials. Diffnet++\footnote{https://github.com/PeiJieSun/diffnet/tree/master/Diffnet\%2B\%2B}, MHCN\footnote{https://github.com/Coder-Yu/RecQ} and NeuralSparse\footnote{https://github.com/flyingdoog/PTDNet/tree/main/NeuralSparse} are implemented based on open source code. It should be noted that NeuralSparse is not an official implementation. The implementation of SocialGCN is based on Diffnet++. We removed the attention module and made some corresponding modifications according to the original paper~\cite{wu2018socialgcn}.

\subsection{Implementation Detail.}
For all experiments, the embedding size and batch size are set to 8 and 1024 respectively, and they all use the Adam optimizer. All social recommendation models use BPR loss. Hyper-parameters like learning rate and dropout are obtained by grid search. The search range is as follows: the learning rate is \{0.005, 0001, 0.0005, 0.0001\}, the dropout is [0, 1], and the weights ($\alpha$) for the co-optimization in GDMSR are \{0.1, 0.3, 0.5, 0.7, 0.9\}. For adaptive denoising strategy, the threshold($\epsilon$) is set to 5, the search range of $\gamma$ and $R$ are \{0.5, 1.0, 2.0\} and \{0.01, 0.015, 0.02, 0.025, 0.03\} respectively. For NeuralSparse, the number of neighbors sampled by each node is searched from \{30, 50, 100\}. For all experiments of GDMSR, we train the denoising model for 200 epochs, at which time $\mathcal{L}^{BCE}$ was basically stable. For all datasets, the interaction history of each user has been preprocessed to have a fixed length, \text{i.e.}, padding for those shorter than $L$ and truncation for excess ones. The truncation follows a descending order based on item popularity in datasets.

\subsection{Loss Curve of Denoising Training}

\begin{figure*}
    \centering
    \subfigure[]{\includegraphics[width=.31\textwidth]{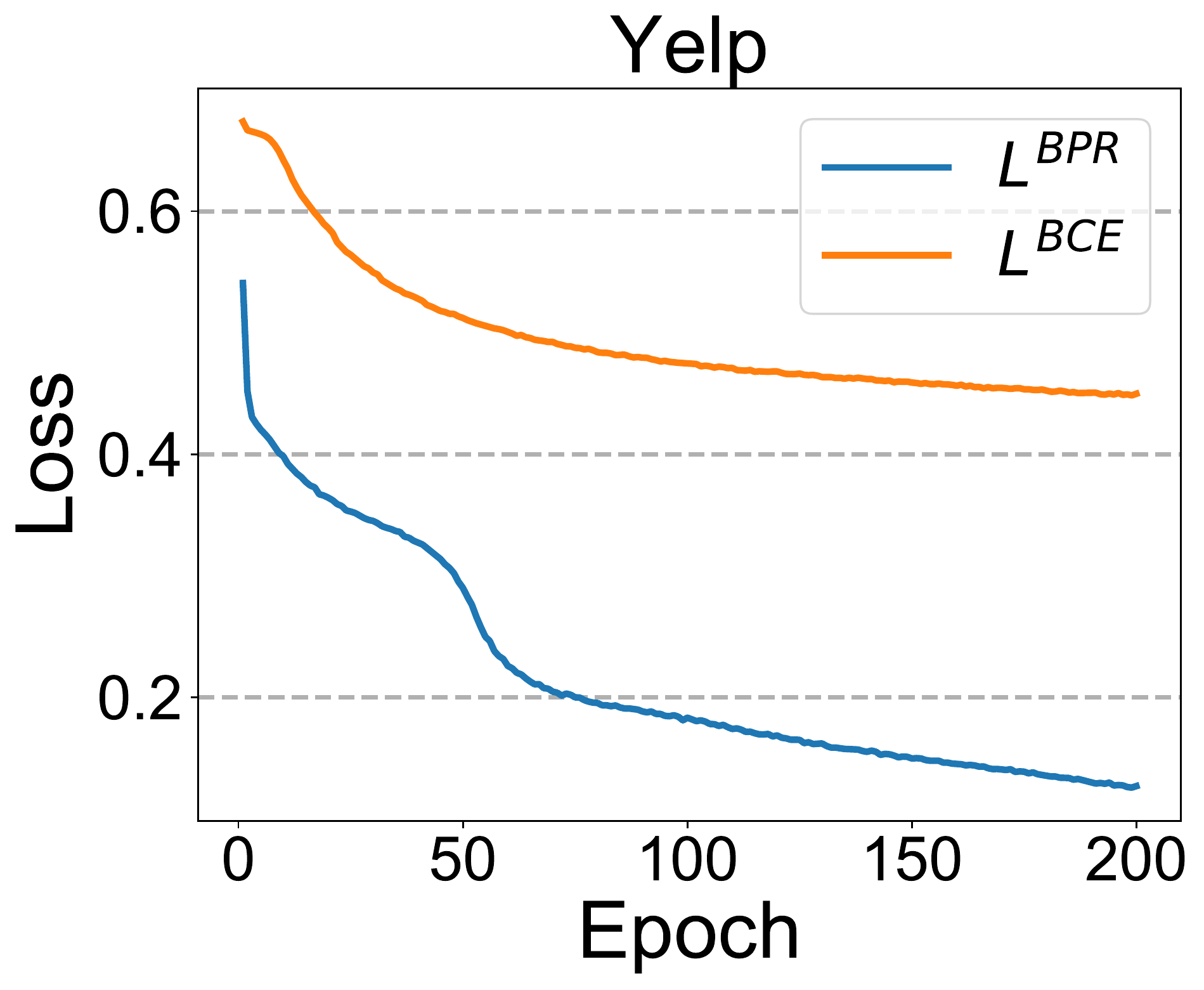}}
    \subfigure[]{\includegraphics[width=.31\textwidth]{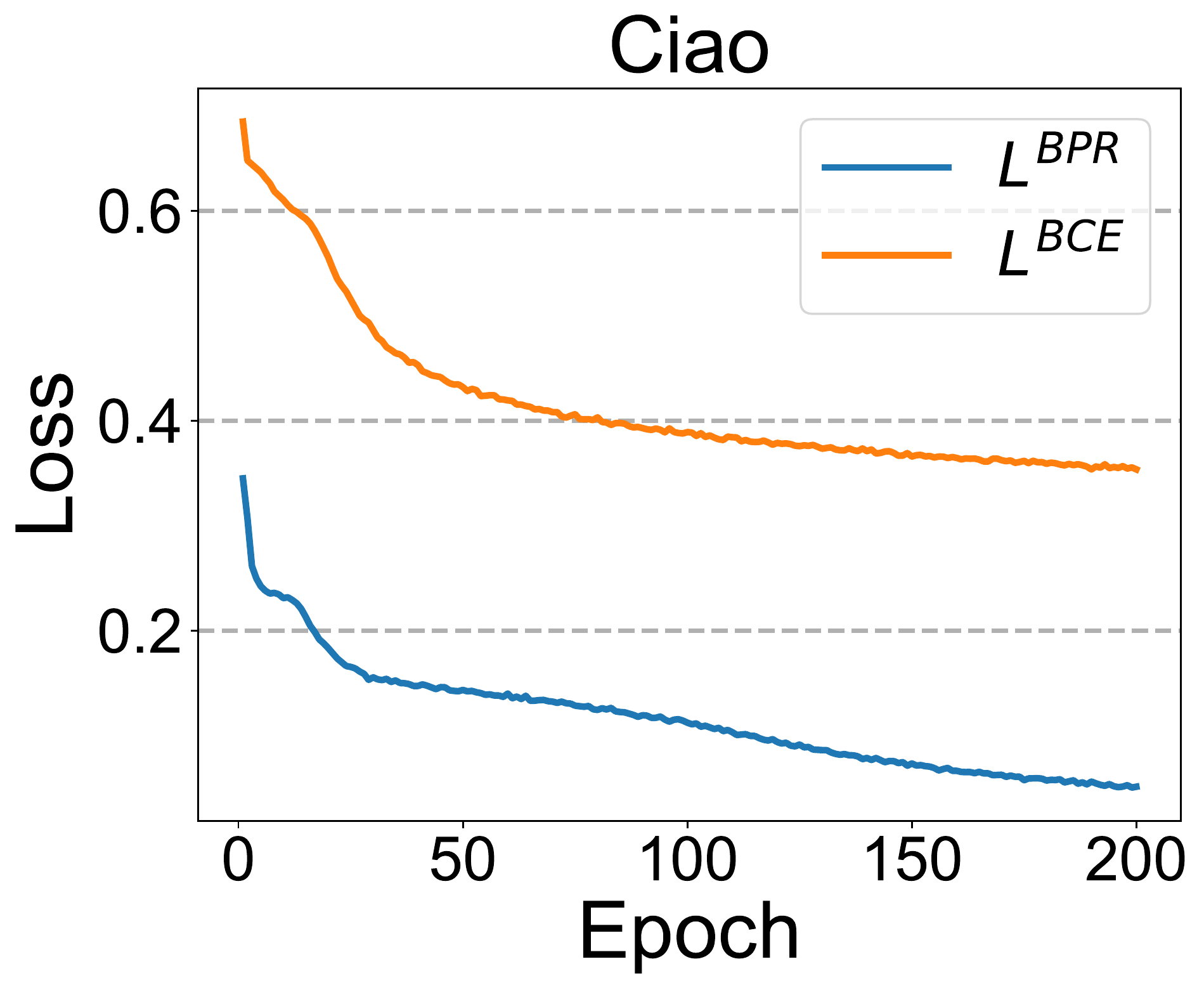}}
    \subfigure[]{\includegraphics[width=.31\textwidth]{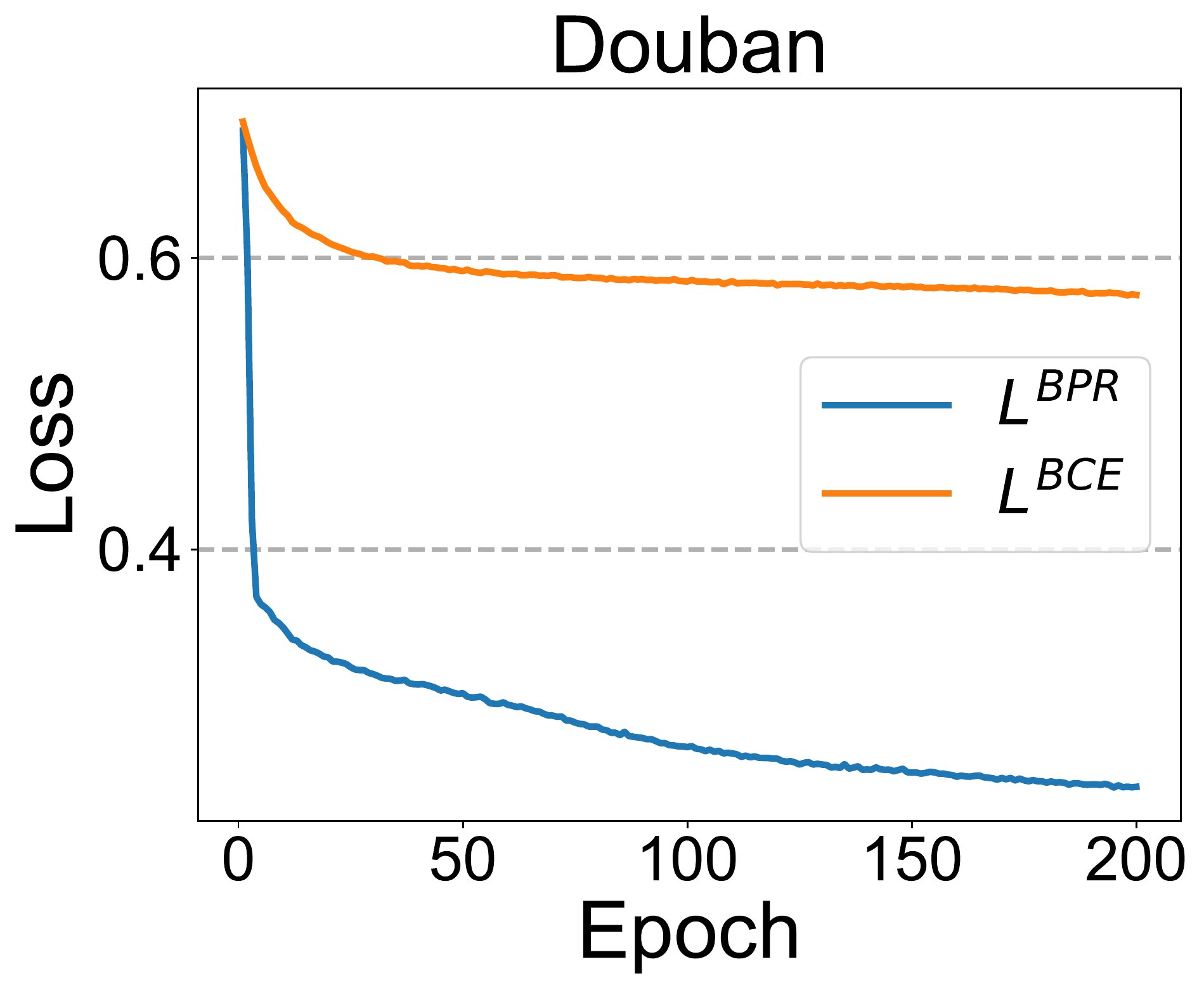}}
    \caption{Loss curve of denoising training}
    \label{fig:loss_curve}
\end{figure*}

In denoising training, for all datasets, we train for 200 epochs so that $L^{BCE}$ is stable, and the loss curves are shown in the Figure~\ref{fig:loss_curve}. We can observe that after 200 epochs of training, the loss function has basically stabilized, indicating that the denoising training has been completed.

\subsection{Experiment details}
In the experiments, we control the number of social relations that are finally removed by controlling different hyper-parameters for GDMSR and baselines, respectively. Specifically, for GDMSR, we control the $\epsilon$, $\gamma$ and $R$. For Rule based approach, we control the ratio of friends removed per user. For NeuralSparse, We control the hyperparameter $K$, which is the maximum number of neighbors that each node can receive message. For ESRF, We also control the number of neighbors each node uses, in the source code, this value is set to 100, and no hyperparameters are provided to tune. We made modifications directly in the source code.

\begin{figure*}
    \centering
    \subfigure[]{\includegraphics[width=.23\textwidth]{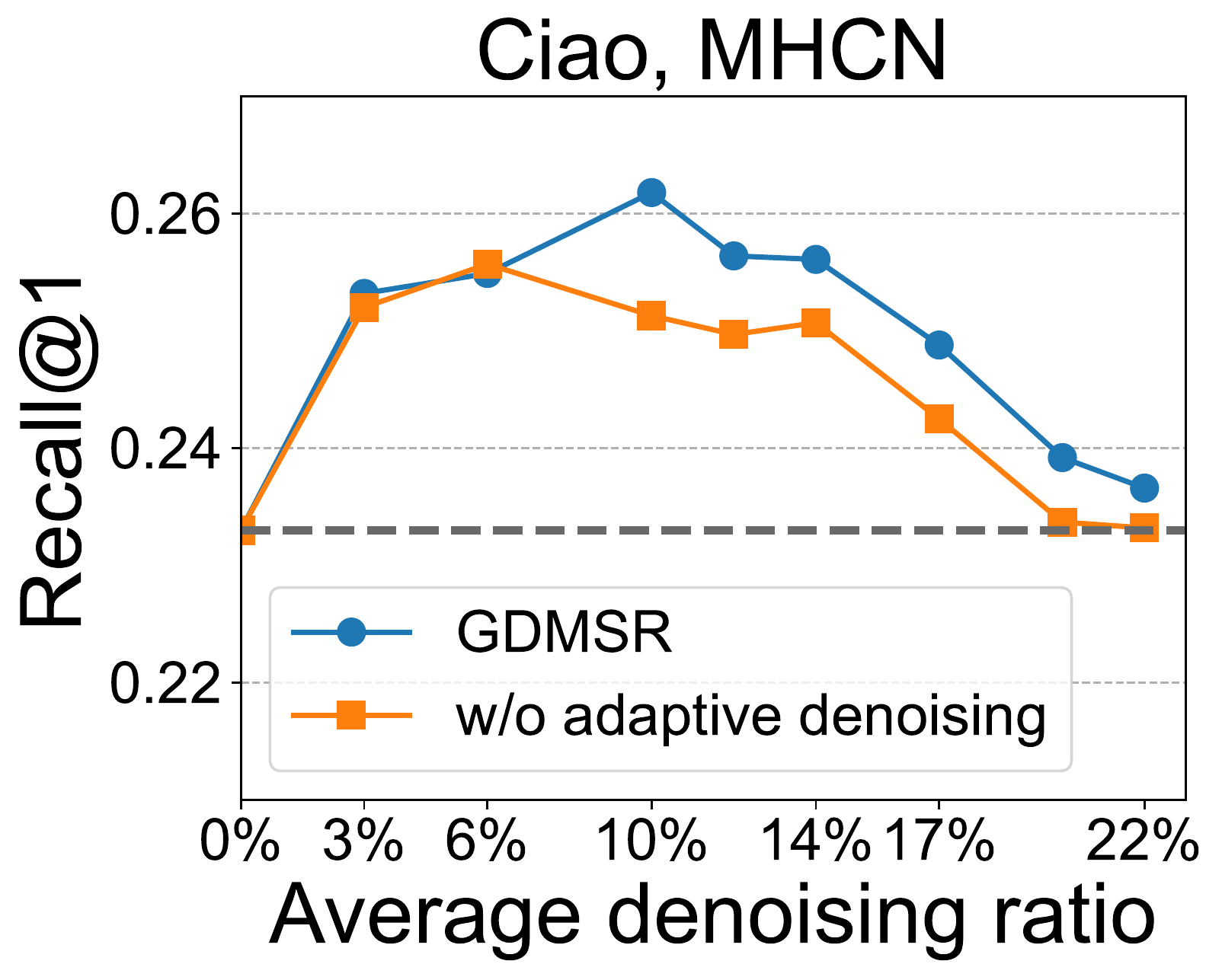}}
    \subfigure[]{\includegraphics[width=.23\textwidth]{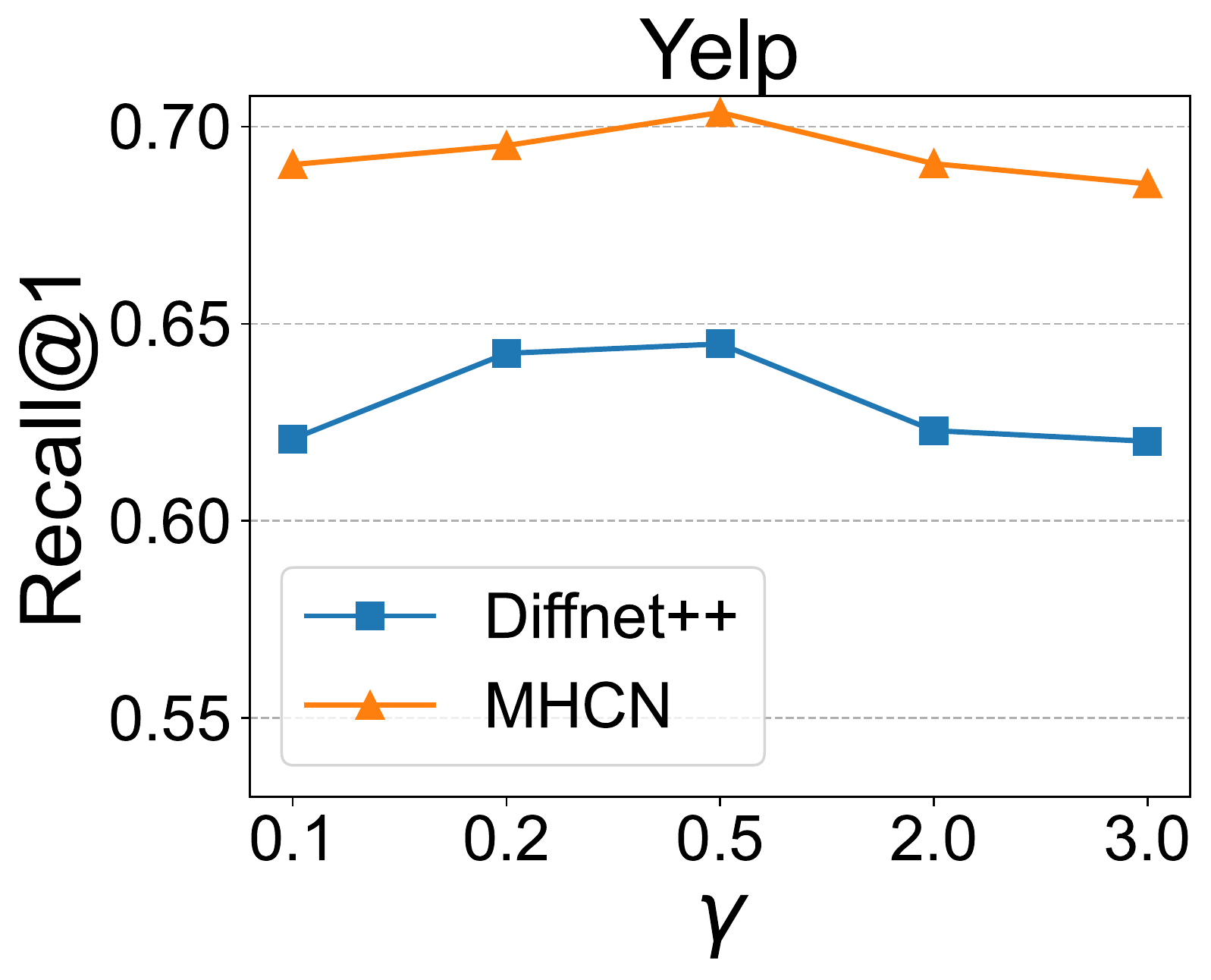}}
    \subfigure[]{\includegraphics[width=.23\textwidth]{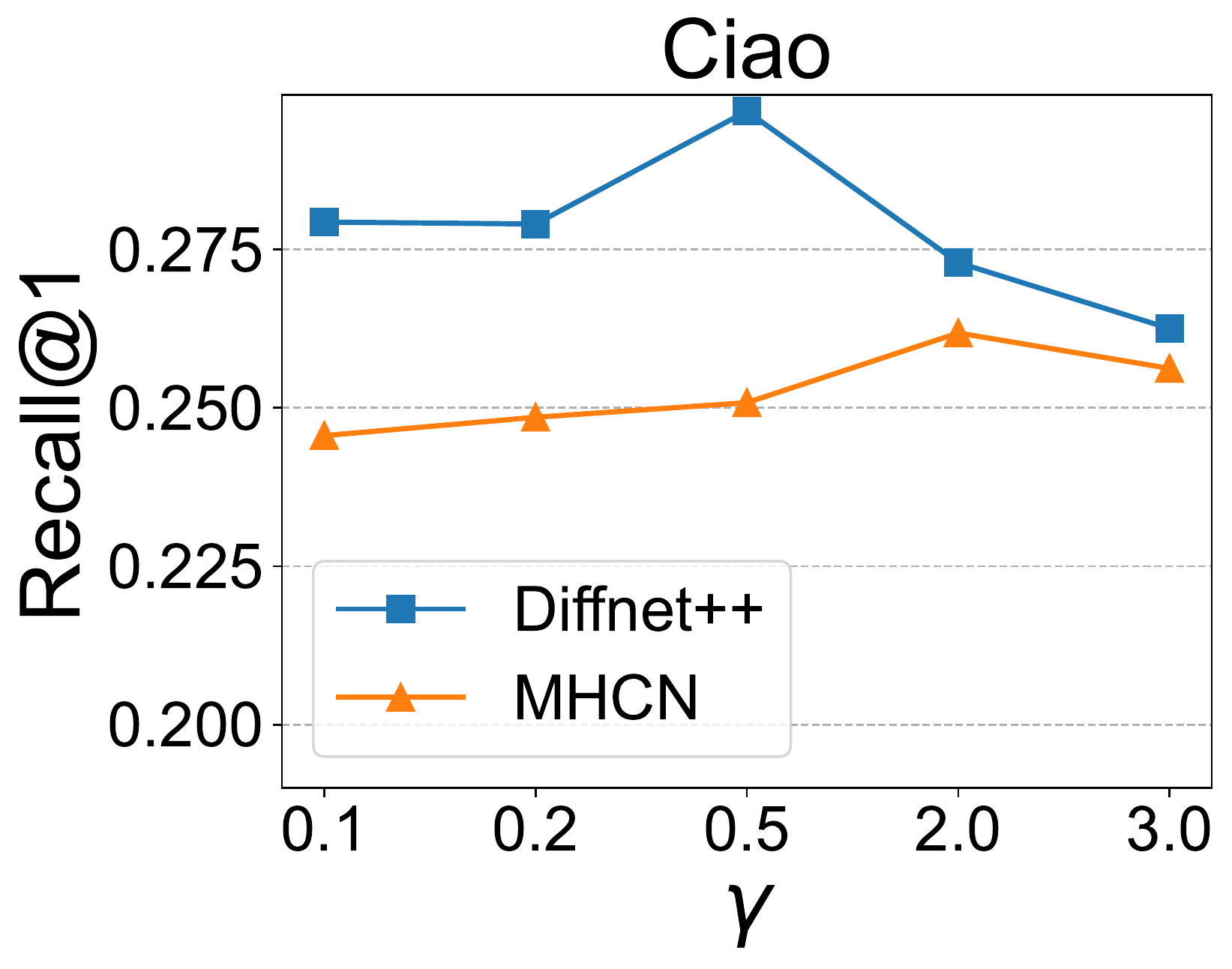}}
    \subfigure[]{\includegraphics[width=.23\textwidth]{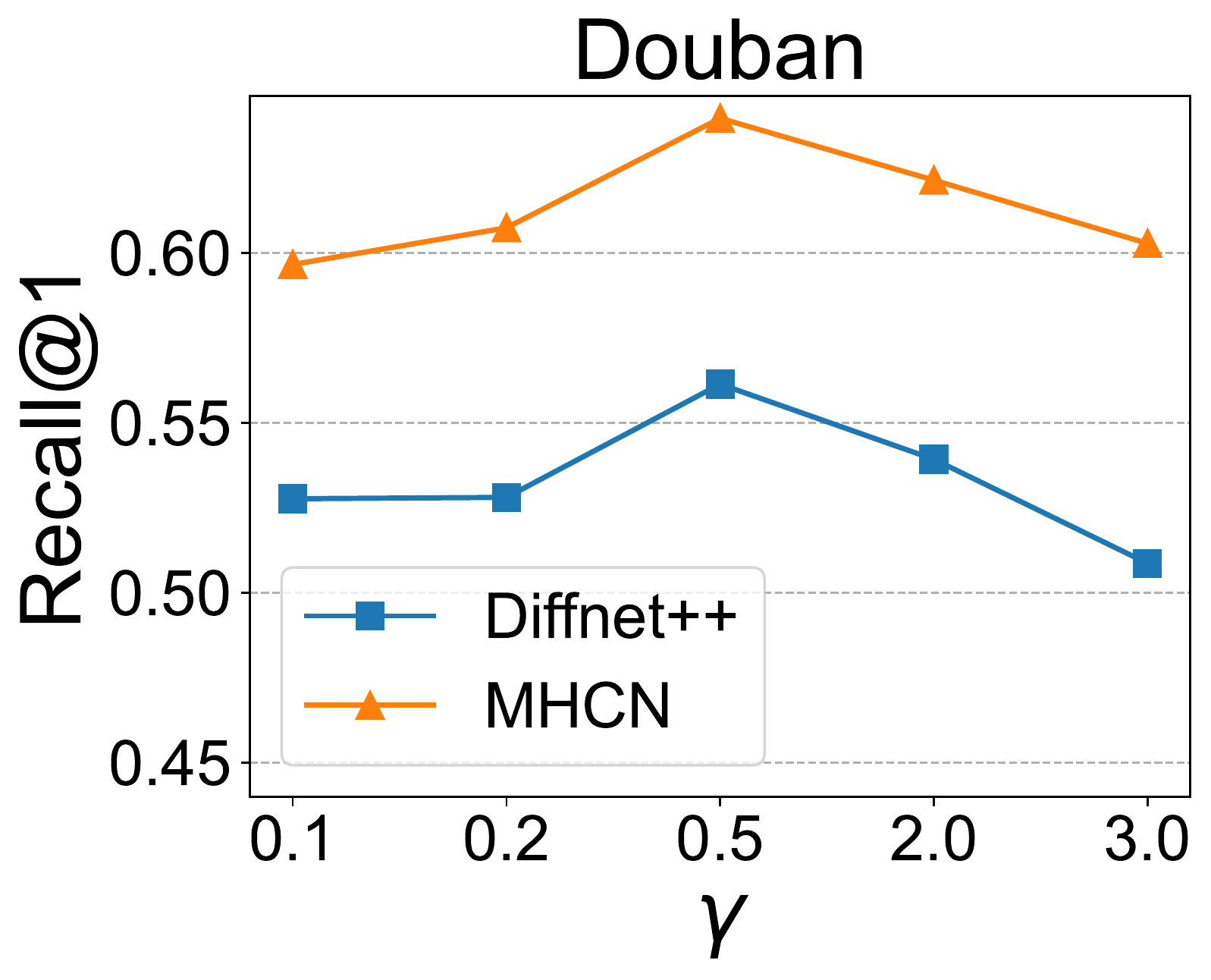}}
    \caption{(a)Performance comparison between different denoising strategy. (b)Performance on different $\gamma$ for adaptive denoising}
    \label{fig:sampling_ablation_appendix}
\end{figure*}

\begin{figure*}
    \centering
    \subfigure{\includegraphics[width=.23\textwidth]{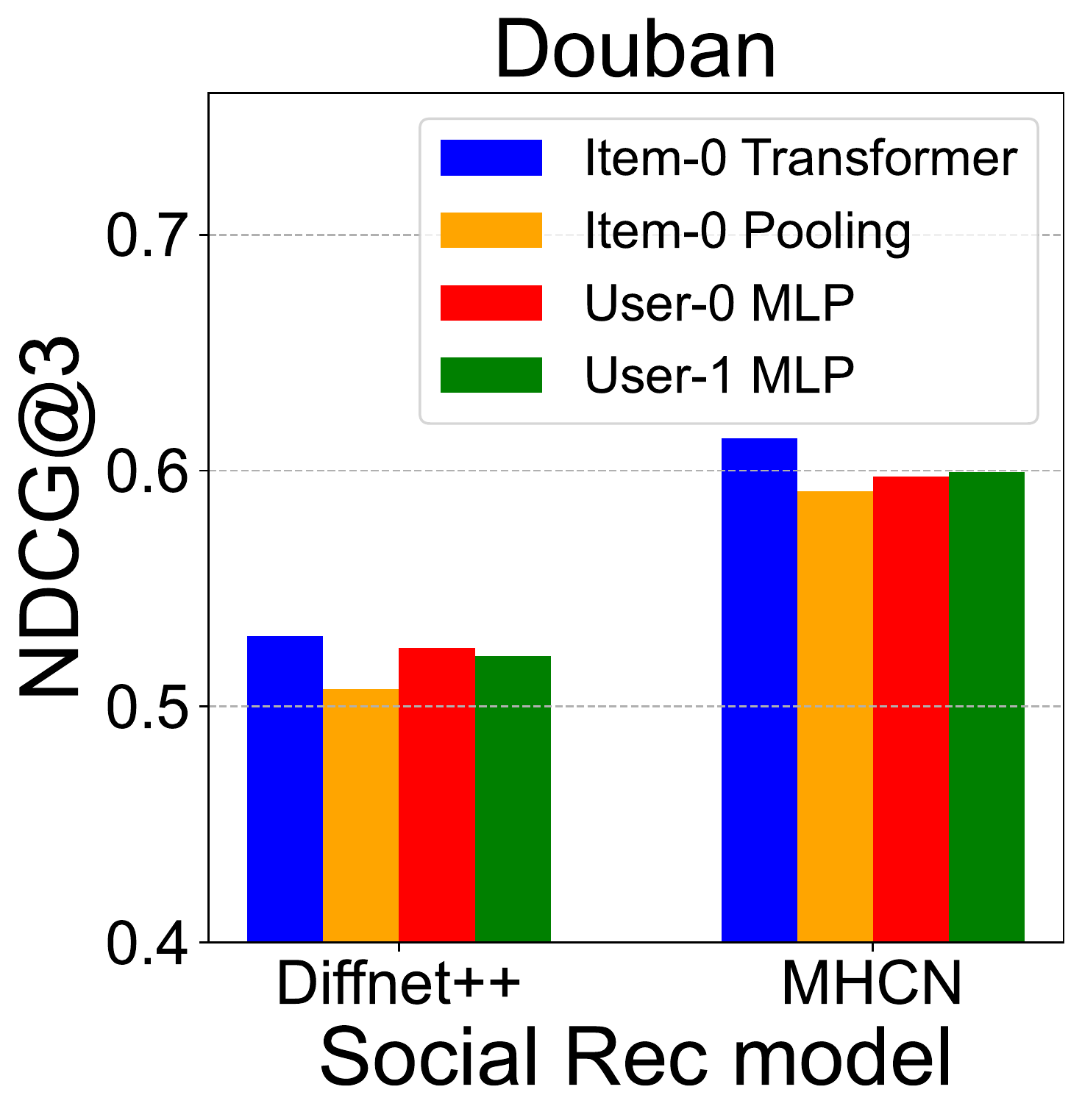}}
    \subfigure{\includegraphics[width=.23\textwidth]{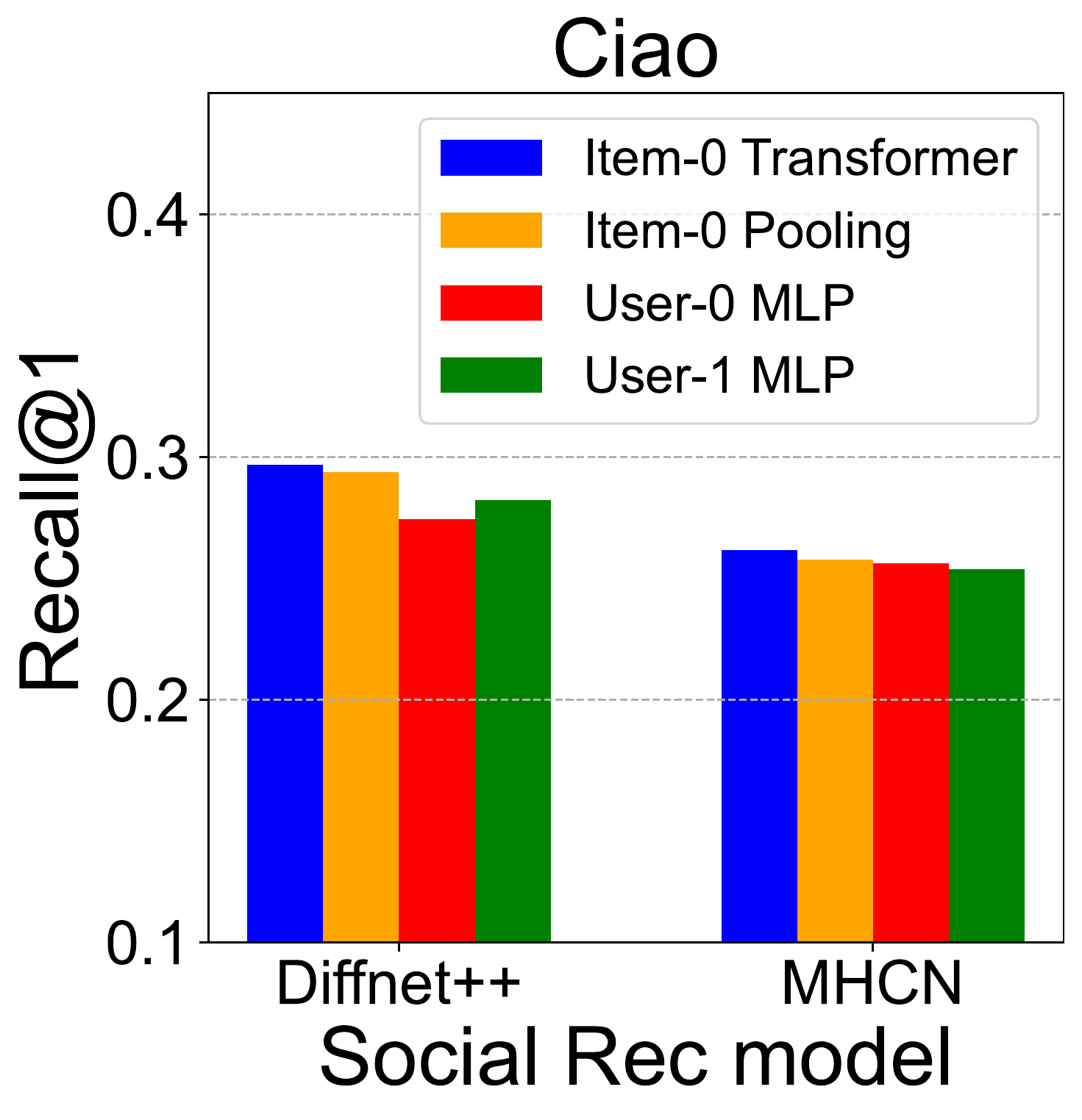}}
    \subfigure{\includegraphics[width=.23\textwidth]{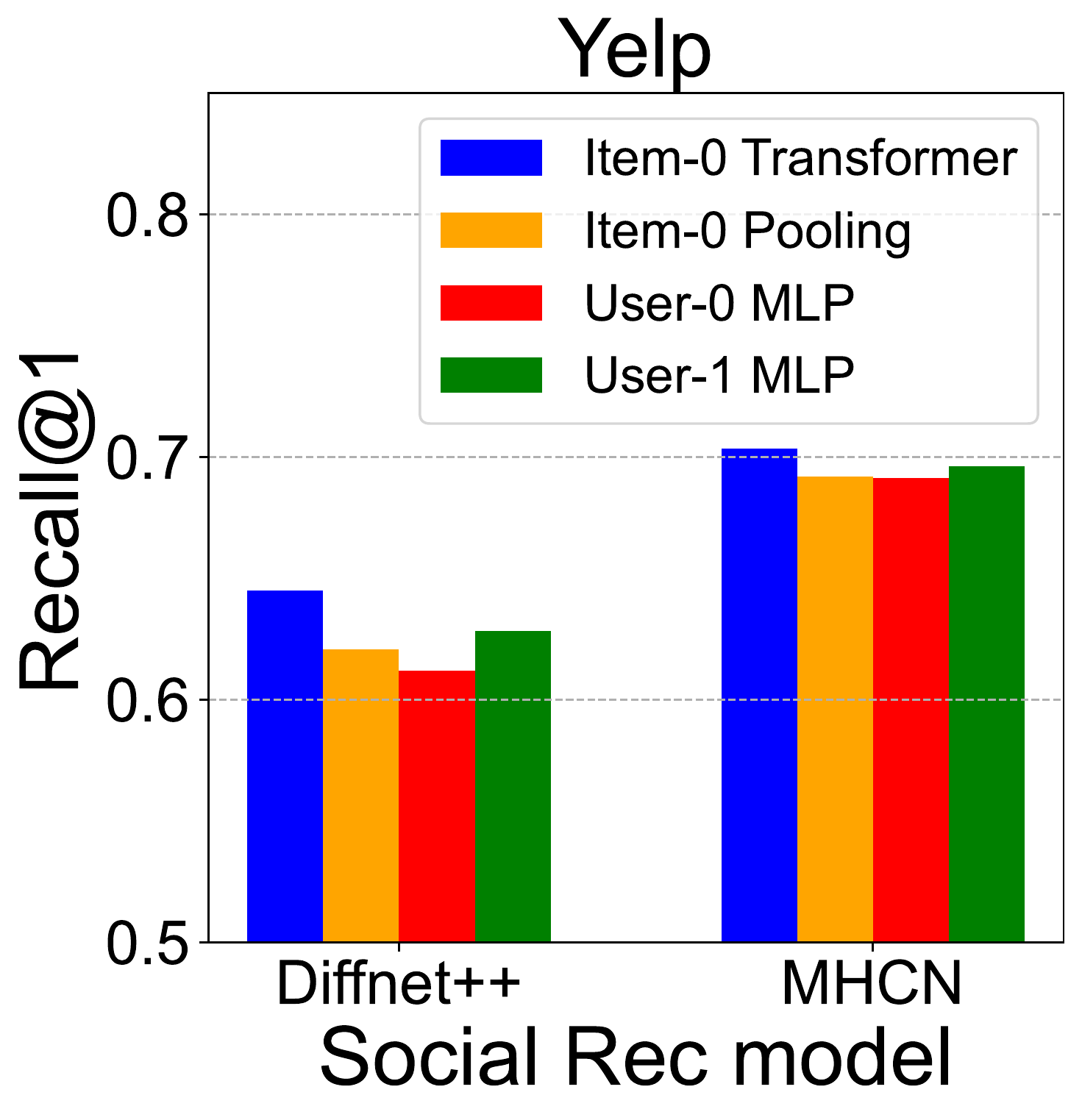}}
    \subfigure{\includegraphics[width=.23\textwidth]{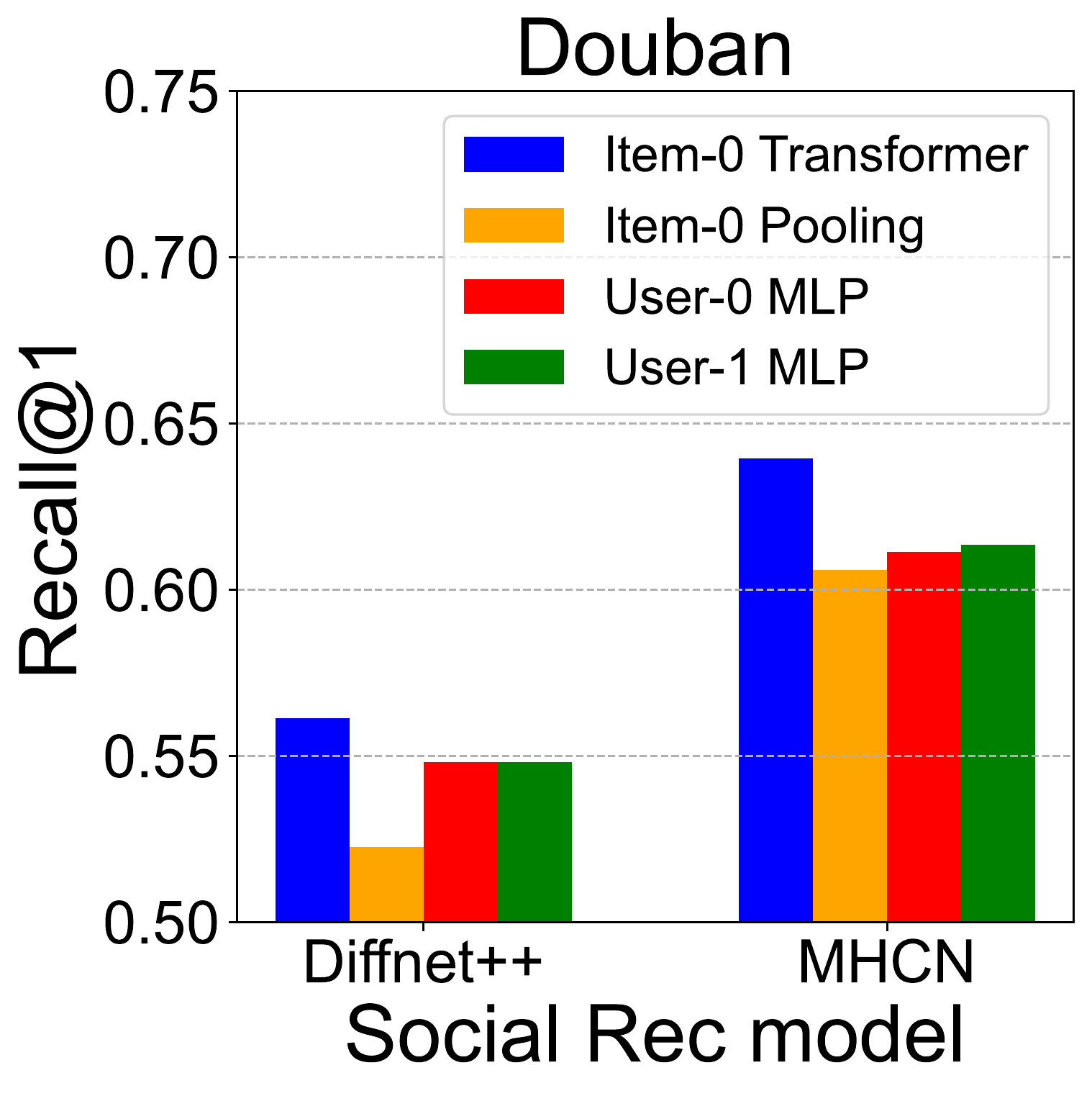}}

    \caption{Performance comparison among different relation confidence modeling structures.}
\end{figure*}

\subsection{Denoising Robustness on Yelp}
\label{sec:appendix_denoising}

To verify the robustness of the adaptive denoising strategy, we compared the strategy to the method that assign same denoising ratio for all users, and we study the effect of different $\gamma$ on the performance. The results are shown in Figure~\ref{fig:sampling_ablation_appendix}(b)-(d). When $\gamma=0.5$, the overall performance is the best. This means that except for those users with very few friends, most users are assigned a larger denoising ratio to achieve better performance.

\end{document}